\documentclass[]{pasj02}
\usepackage[switch,mathlines]{lineno} % add line number to manuscript

\jyear{2025}
\Received{2025/04/01}%{yyyy/mm/dd}
\Accepted{2025/05/12}%{yyyy/mm/dd}
\begin{document}
    
%%----------------------------------------------------------------------------
\title{Thermal and Kinematic Properties of Ejecta in SN1987A revealed by XRISM}
%%----------------------------------------------------------------------------

%%% begin:list of authors
% Do NOT capitalize all letters in "textsc".
\author{XRISM Collaboration}
%\thanks{Corresponding Authors: Yukikatsu Terada, Koji Mori, Tsukasa Matsushima, Satoru Katsuda, and Aya Bamba}
\footnotetext[*]{Corresponding authors: Yukikatsu Terada, Koji Mori, Tsukasa Matsushima, Satoru Katsuda, and Aya Bamba.}
%\footnotetext[*]{Corresponding Authors.}
\footnotetext[$\dag$]{E-mail: terada@mail.saitama-u.ac.jp}
%\email{terada@mail.saitama-u.ac.jp}

%%% XRISM Science Member +XGS invited
\author{Marc \textsc{Audard}\altaffilmark{1}\orcid{0000-0003-4721-034X}}
\author{Hisamitsu \textsc{Awaki}\altaffilmark{2}}
\author{Ralf \textsc{Ballhausen}\altaffilmark{3,4,5}\orcid{0000-0002-1118-8470}}
\author{Aya \textsc{Bamba}\altaffilmark{6,*}\orcid{0000-0003-0890-4920}}
\author{Ehud \textsc{Behar}\altaffilmark{7}\orcid{0000-0001-9735-4873}}
\author{Rozenn \textsc{Boissay-Malaquin}\altaffilmark{8,4,5}\orcid{0000-0003-2704-599X}}
\author{Laura \textsc{Brenneman}\altaffilmark{9}\orcid{0000-0003-2663-1954}}
\author{Gregory V.\ \textsc{Brown}\altaffilmark{10}\orcid{0000-0001-6338-9445}}
\author{Lia \textsc{Corrales}\altaffilmark{11}\orcid{0000-0002-5466-3817}}
\author{Elisa \textsc{Costantini}\altaffilmark{12}\orcid{0000-0001-8470-749X}}
% ...

%%% Affil
\altaffiltext{1}{Department of Astronomy, University of Geneva, Versoix CH-1290, Switzerland} %Geneva U
\altaffiltext{2}{Department of Physics, Ehime University, Ehime 790-8577, Japan} %Ehime U
\altaffiltext{3}{Department of Astronomy, University of Maryland, College Park, MD 20742, USA} %U of Maryland
\altaffiltext{4}{NASA / Goddard Space Flight Center, Greenbelt, MD 20771, USA} %NASA/GSFC
\altaffiltext{5}{Center for Research and Exploration in Space Science and Technology, NASA / GSFC (CRESST II), Greenbelt, MD 20771, USA} 
\altaffiltext{6}{Department of Physics, University of Tokyo, Tokyo 113-0033, Japan} % U of Tokyo
\altaffiltext{7}{Department of Physics, Technion, Technion City, Haifa 3200003, Israel}
\altaffiltext{8}{Center for Space Science and Technology, University of Maryland, Baltimore County (UMBC), Baltimore, MD 21250, USA}
\altaffiltext{9}{Center for Astrophysics | Harvard-Smithsonian, MA 02138, USA} %CfA
\altaffiltext{10}{Lawrence Livermore National Laboratory, CA 94550, USA} %LLNL
\altaffiltext{11}{Department of Astronomy, University of Michigan, MI 48109, USA} %U of Michigan
\altaffiltext{12}{SRON Netherlands Institute for Space Research, Leiden, The Netherlands} %SRON
% ...

%% !!! Select 3 to 5 words from PASJ's key words !!! 
%% List of Key Words: https://academic.oup.com/pasj/pages/Pasj_Keywords 
%% "\KeyWords{ }" always has to be placed before ``\maketitle'' 
\KeyWords{supernovae: individual (SN1987A) --- ISM: supernova remnants --- plasmas --- shock waves}  

\maketitle

%================== Abstract =====================================
\begin{abstract}
We present an analysis of high-resolution spectra from the shock-heated plasmas in SN~1987A, based on an observation using the Resolve instrument onboard the X-Ray Imaging and Spectroscopy Mission (XRISM). 
The 1.7--10 keV Resolve spectra are accurately represented by a single component, plane-parallel shock plasma model, with a temperature of $2.84_{-0.08}^{+0.09}$ keV and an ionization parameter of $2.64_{-0.45}^{+0.58} \times 10^{11}$ s cm$^{-3}$. 
The Resolve spectra are also well reproduced by the 3-D magneto-hydrodynamic simulation presented by \citet{2020A&A...636A..22O} suggesting substantial contribution from the ejecta.
The metal abundances obtained with Resolve align with the LMC value, indicating that the X-rays in 2024 originate from non-metal-rich shock-heated ejecta and the reverse shock has not reached the inner metal-rich region of ejecta.
Doppler widths of the atomic lines from Si, S, and Fe correspond to velocities of 1,500--1,700 km s$^{-1}$, where the thermal broadening effects in this non-metal-rich plasma are negligible.
Therefore, the line broadening seen in Resolve spectra is determined by the large bulk motion of ejecta.
For reference, we determined a 90\% upper limit on non-thermal emission from a pulsar wind nebula at $4.3 \times 10^{-13}$ erg cm$^{-2}$ s$^{-1}$ in the 2 -- 10 keV range, aligning with NuSTAR findings by \citet{2022ApJ...931..132G}. 
Additionally, we searched for the $^{44}$Sc K line feature and found a $1\sigma$ upper limit of $1.0 \times 10^{-6}$ photons cm$^{-2}$ s$^{-1}$, which translates to an initial $^{44}$Ti mass of approximately $2 \times 10^{-4} M_{\odot}$, consistent with previous X-ray to soft gamma-ray observations \citep{2015Sci...348..670B,2012Natur.490..373G,2006ApJ...651.1019L}.
\end{abstract}
%================== Abstract =====================================

%\pagewiselinenumbers 

%=================================================================
\section{Introduction}
\label{sec:introduction}
%=================================================================

%%%% Overview
Supernovae and their remnants release heavy elements and cosmic rays into space, thus contributing to the diversity of the universe.
X-ray observations are crucial to understanding their kinematics and the degree of enrichment of heavy elements they contribute.
%%%% SN1987A
SN~1987A in the Large Magellanic Cloud (LMC) at a distance of 51.4 kpc is a core-collapse supernova that was observed to explode on 1987 February 23 \citep{1987A&A...177L...1W}.
It is the only supernova that has been monitored by human beings for nearly 40 years as it transitions into a supernova remnant. 
The object is evolving in a highly inhomogeneous circumstellar medium (CSM), which was originally created by stellar winds from the massive progenitor. 
Early optical observations using the Hubble Space Telescope identified a triple-ring structure, featuring a luminous inner equatorial ring (ER) and two dim outer rings \citep{1995ApJ...452..680B}. 
The CSM is characterized by a dense, clumpy ER within a more diffuse H{\sc ii} region \citep{1995ApJ...452L..45C,2005ApJS..159...60S}. 
%%%% X-ray observations
The blast wave generated by the explosion continues to interact with these CSM structures.
As a result, the electrons, protons, and ions in these regions are expected to be heated by the shock, leading to X-ray emission under different temperatures depending on the mass of the species. 
These species will gradually reach equilibrium through Coulomb collisions and collective plasma interactions, although non-equilibrium conditions are expected to last well after the interaction with the shock due to the slow nature of these processes.
The X-rays from hot plasma in SN~1987A have been monitored by numerous X-ray missions, including ROSAT, Suzaku, Chandra, XMM-Newton, NuSTAR, SRG/eROSITA (e.g., \cite{1994A&A...281L..45B,1996A&A...312L...9H,2000ApJ...543L.149B,2002ApJ...567..314P,2004ApJ...610..275P,2005ApJ...628L.127Z,2005ApJ...634L..73P,2006ApJ...645..293Z,2006ApJ...646.1001P,2008ApJ...676L.131D,2009ApJ...703.1752R,2009ApJ...692.1190Z,2009PASJ...61..895S,2010A&A...515A...5S,2010MNRAS.407.1157Z,2011ApJ...733L..35P,2012A&A...548L...3M,2013ApJ...764...11H,2016ApJ...829...40F,2021ApJ...916...41S,2021ApJ...922..140R,2022A&A...661A..30M,2024ApJ...966..147R,2025ApJ...981...26S} and references therein), investigating the shock interactions primarily with the ER but also with the outer envelope of ejecta, exposing the progenitor's evolution history.

%%%% Time evolution, Ejecta dominant.
The X-ray light curve and spectra have been accurately replicated through 3D hydrodynamic and magneto-hydrodynamic (MHD) simulations by \citet{2015ApJ...810..168O}, \citet{2019A&A...622A..73O}, and \citet{2020A&A...636A..22O}. 
Initially, these studies showed that X-ray emission arose primarily from the shocked H{\sc ii} region. 
Over time, the emission has become dominated by the shocked dense ER. 
The model depicts a third stage starting approximately 33 years post-explosion, suggesting that the blast wave is leaving the ER and the shock-heated emission from the ejecta is rising \citep{2020A&A...636A..22O}. 

%%% XRISM and expectation
As of September 2023, X-ray high-resolution spectroscopy within the Fe-K band has been made feasible by the X-Ray Imaging and Spectroscopy Mission (XRISM; \cite{2024SPIE130931G_XRISM}). 
This mission is managed by the Japan Aerospace Exploration Agency (JAXA) and NASA, with collaboration from the European Space Agency (ESA) and other international partners.
Prior to the performance verification (PV) observation of XRISM, \citet{2024ApJ...961L...9S} and \citet{2024RNAAS...8..156S} investigated the plasma diagnostics of SN~1987A, predicting what XRISM's high-resolution spectra would look like in 2024, during the PV phase. 
They performed this study under the gate valve (GV) open and closed conditions, utilizing the latest outputs from the 3D MHD simulation by \citet{2020A&A...636A..22O}. 
They anticipated a strong link between the broadened line emission and newly shocked ejecta.
%%% Paper structure

This paper presents the initial results from the XRISM observation of SN~1987A during the PV phase. 
The organization of the paper is as follows: Sections \ref{sec:observation} and \ref{sec:data_reduction_validation} provide an overview of the XRISM observation of SN~1987A and outline the data reduction process, respectively. Sections \ref{sec:data_analyses_spectrum} and \ref{sec:data_analyses_line} focus on the analysis of the X-ray spectral shape and plasma diagnostics using atomic lines, respectively. 
In Section \ref{sec:discussion}, we explore the origin of X-ray emission observed by XRISM, discuss line diagnostic results, and cover additional topics.

%=================================================================
\section{Observatory and the Observation}
\label{sec:observation}
%=================================================================

%%% XRISM satellite
In order to perform plasma diagnostics of SN~1987A in the X-ray band, we used the X-Ray Imaging and Spectroscopy Mission (XRISM; \cite{2024SPIE130931G_XRISM}) which provides high-resolution X-ray spectroscopy, especially on the Fe-K band at around 6 keV, for the first time.
XRISM carries an X-ray microcalorimeter array (known as Resolve; \cite{2024SPIE_Kelley}) and an X-ray CCD camera (known as Xtend; \cite{2024SPIE_Mori}, \cite{2025arXiv250208030N}) in the focal planes of its two identical X-ray mirrors. 
These instruments were designed to perform X-ray spectroscopy with energy resolution of $\le7$~eV (Full Width at Half Maximum; FWHM) in a field of view (FOV) of $3.05 \times 3.05$ arcmin$^2$ and X-ray imaging over a wide FOV of $38.5 \times 38.5$ arcmin$^2$ in the 0.3--12~keV and 0.4--13~keV bands, respectively.

%%% SN1987A observation
We have performed an X-ray observation of SN~1987A with Resolve and Xtend onboard XRISM, starting at 18:20:04 on 2024 June 17 until 14:01:04 2024 June 22 UT during the performance verification and calibration period in the nominal operations phase.
This observation took place 37.3 years after the observed explosion of SN~1987A.
The observation id of this observation was OBSID = 300021010.
The target position is (RA, Dec)$_{\rm J2000.0}$ = (\timeform{83D866744}, \timeform{-69D268637}) with the roll angle of \timeform{31D012931}.
The Resolve instrument was operated with no filter, but in gate valve (GV) closed condition, which limits the energy band pass of Resolve to 1.7 -- 10 keV.
Xtend was operated in nominal mode with the full window option.

%=================================================================
\section{Data Reduction and Validation}
\label{sec:data_reduction_validation}
%=================================================================
%------------------------------------------------
\subsection{Data Reduction}
\label{sec:data_reduction_validation:process}
%------------------------------------------------
The X-ray data acquired via the XRISM satellite were processed by the standard XRISM pre-pipeline and pipeline \citep{2020SPIE11444E..5DL,2021JATIS...7c7001T,2024SPIE_Hayashi} with the version of TLM2FITS= '005\_001.20Jun2024\_Build8.012' and PROCVER = '03.00.013.009', respectively.
We employed XRISM ftools available in the HEAsoft package version 6.34 together with the CALDB XRISM version 10 for our analyses.

In order to perform detailed plasma diagnostics of hot plasma within SN~1987A through high-resolution spectroscopy, our study concentrates on the Resolve instrument; future publications will present analyses involving Xtend.
The standard cleaned events of Resolve were derived using the nominal screening criteria applied in the pipeline process plus the additional screening, advised in Section 6 of the XRISM quick start guide version 2.3 dated 18 September 2024, was carried out via the RISE\_TIME and DERIV\_MAX parameters as follows: {\tt ((PI>=600) \&\& ((RISE\_TIME+0.00075*DERIV\_MAX)>46) \&\& ((RISE\_TIME+0.00075*DERIV\_MAX)<58))}.
We used the Resolve events with the High primary (Hp) grade \citep{2024SPIE_Kelley} in the following analyses.
The total exposure for Resolve was 290.5 ksec.

%============== Figure 1 ====================
\begin{figure}[htb]
    \begin{center}
    \includegraphics[width=0.48 \textwidth]{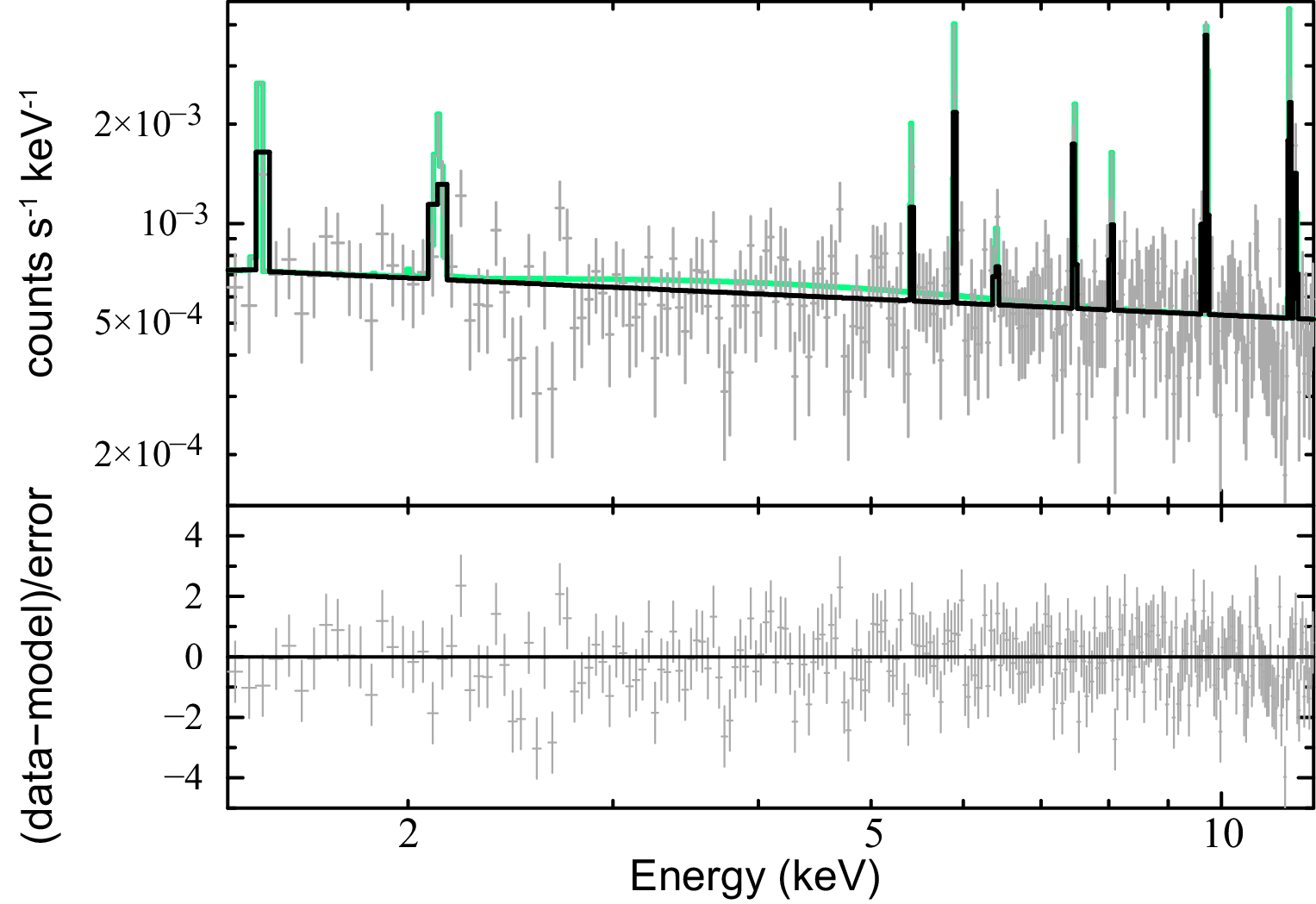}
    \end{center}
    \caption{The NXB spectra are depicted as gray crosses, while the model is represented by a black line, with Resolve.
    The contamination of LMC emission on Resolve estimated by the Xtend background region is shown in green. 
    {Alt text: Single line graph illustrating both the data and the models.} 
     }
    \label{fig:background_spectra_resolve_xtend}
\end{figure}
%============== Figure 1 ===================

%% Resolve gain
The gain drift of Resolve was tracked using the Mn-K lines from an $^{55}$Fe source on the filter wheel onboard the spacecraft, which was intermittently exposed to all the pixels.
As a cross-check, the behavior of the gain of the calibration pixel \#12 is always illuminated by Mn-K lines.
After the standard gain correction, the energy resolutions of the former and latter events were $4.49 \pm 0.02$ eV for Hp (variation from 4.2 -- 5.0 eV by pixel) and $4.50 \pm 0.01$ eV, respectively.
The performance of the energy resolution of Resolve pixels during the SN~1987A observation exceeds the mission requirement.
Therefore, we used all the pixels except for the calibration pixel \#12 for the following spectral analyses, although pixel 27 exhibits a behavior in slightly different gain trending from other pixels.
The total count rate for Hp cleaned events in the 1.7 -- 10 keV band is $5.9 \times 10^{-2}$ counts s$^{-1}$.

%------------------------------------------------
\subsection{Energy response and Spectral Model}
\label{sec:data_reduction_validation:spectrum}
%------------------------------------------------
%%% Response function
In the following spectral analyses, we used XSPEC version 12.14.1 in HEAsoft package version 6.34. 
We used the extra-large sized response function with electron-loss continuum component generated by the ftool {\it rslmkrmf} and the point source arf file generated by the ftool {\it xaarfgen} at the position of SN1987A for all pixels. 
%%% Binning, cash statistics
The Resolve spectrum has been binned in an optimal manner following the method outlined by \citet{2016A&A...587A.151K}. 
The spectral fitting in this article was conducted utilizing C-statistics (C-stat; \cite{1979ApJ...228..939C}).

%%% Models
We applied the models {\it wabs*phabs*zashift} in XSPEC, multiplying to (and convolving) the trial models in the following analyses. 
This approach accounted for the Galactic absorption of $6.0\times 10^{20}$ atoms cm$^{-2}$ (in {\it wabs}) using the photoelectric cross section from \citet{1983ApJ...270..119M}, incorporated the photo absorption in LMC with the column density of $2.2\times 10^{21}$ atoms cm$^{-2}$ (in {\it phabs}) using the LMC abundance taken from \citet{1992ApJ...384..508R} (and \cite{2016AJ....152..110S} for Si), and adjusted the photon energies for the recession velocity of SN~1987A at 286.7 km s$^{-1}$ (\citet{1987ICRC....1..164C,1995Ap&SS.233...75M}; in {\it zashift}).
We also applied the LMC abundance by \citet{1992ApJ...384..508R} in the plasma models for SN~1987A, such as {\it apec} \citep{2001ApJ...556L..91S}, {\it pshock} model families, etc.
We fixed all the parameters of {\it wabs*phabs*zashift} here in the following analyses. 
We used the AtomDB version 3.1.2 for {\it apec} family models.

%------------------------------------------------
\subsection{Background spectra}
\label{sec:data_reduction_validation:background}
%------------------------------------------------
%%% NXB
The non X-ray background (NXB) spectrum was generated by the ftool {\it rslnxbgen} with the provisional version 1 of the NXB database \footnote{https://heasarc.gsfc.nasa.gov/docs/xrism/analysis/nxb/index.html}, employing the same selection criteria as for the cleaned events described in Section \ref{sec:data_reduction_validation:process}. 
The NXB spectral model, which is also provided by the XRISM collaboration in this provisional version, reproduced this NXB spectrum for this observation as shown in Figure \ref{fig:background_spectra_resolve_xtend}. The C-stat was 16347.28 for 21992 degrees of freedom (dof) and the scale factor relative to the nominal NXB model was $0.96 \pm 0.02$, where the uncertainty signifies a 90\% confidence level.

%%% BGD LMC
In order to estimate the contamination from the LMC emission around SN~1987A in Resolve spectra, we analyzed the Xtend data. 
The background Xtend spectrum was derived from a $200" \times 200"$ source-free region in the Xtend image. 
The Xtend NXB spectrum was produced using the ftool {\it xtdnxbgen}, employing version 1 of the NXB database. 
The background Xtend spectrum showed enhancement over the NXB spectrum in the soft energy range below 2 keV and was accurately modeled using the two-temperature optically-thin thermal plasma model (i.e., {\it vapec} in XSPEC; \citet{2001ApJ...556L..91S}), achieving a C-stat of 191.28 with 190 dof, utilizing a flat arf for the background area. 
This emission model indicates that the contamination from the soft background emission in Resolve's spectrum is approximately $2 \times 10^{-4}$ counts s$^{-1}$, constituting 0.3\% of the source signal.
This value is negligible as indicated in Figure \ref{fig:background_spectra_resolve_xtend}, and therefore, for the following analyses, only the NXB spectrum was applied as a background for the Resolve spectra.

%%% Solare flare activity
During the observation of SN~1987A with XRISM, nine M-class solar flares were reported \footnote{https://swc.nict.go.jp/en/report/view.html}, which might enhance the background spectra. 
In the Xtend data, we identified hard X-ray emission above 2 keV from a source-free image region, showing a 0.8 count s$^{-1}$ rate in the 2 -- 10 keV range over 12-minute intervals around 13:00 on each of 18 and 19 June 2024 UT. 
The possible contamination affecting the Resolve spectra is only 1 count in the 2 -- 10 keV band, which is less than 0.01\% of the SN~1987A signals. 
As such, we ignored solar flare contamination on the Resolve spectra in subsequent analyses.

%============== Figure 2 ====================
\begin{figure}[thb]
    \begin{center}
    \includegraphics[width=0.47 \textwidth]{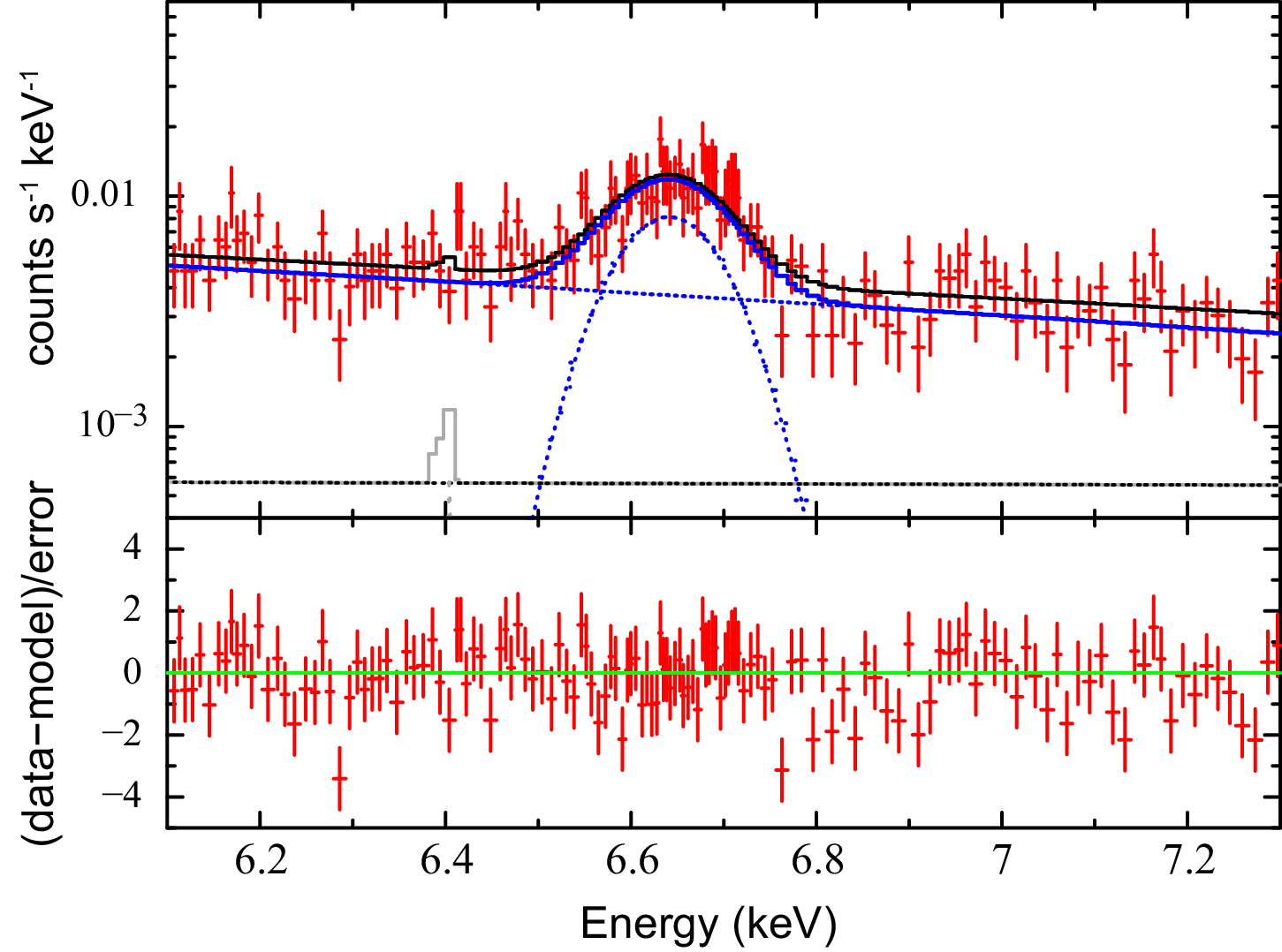}
    \includegraphics[width=0.47 \textwidth]{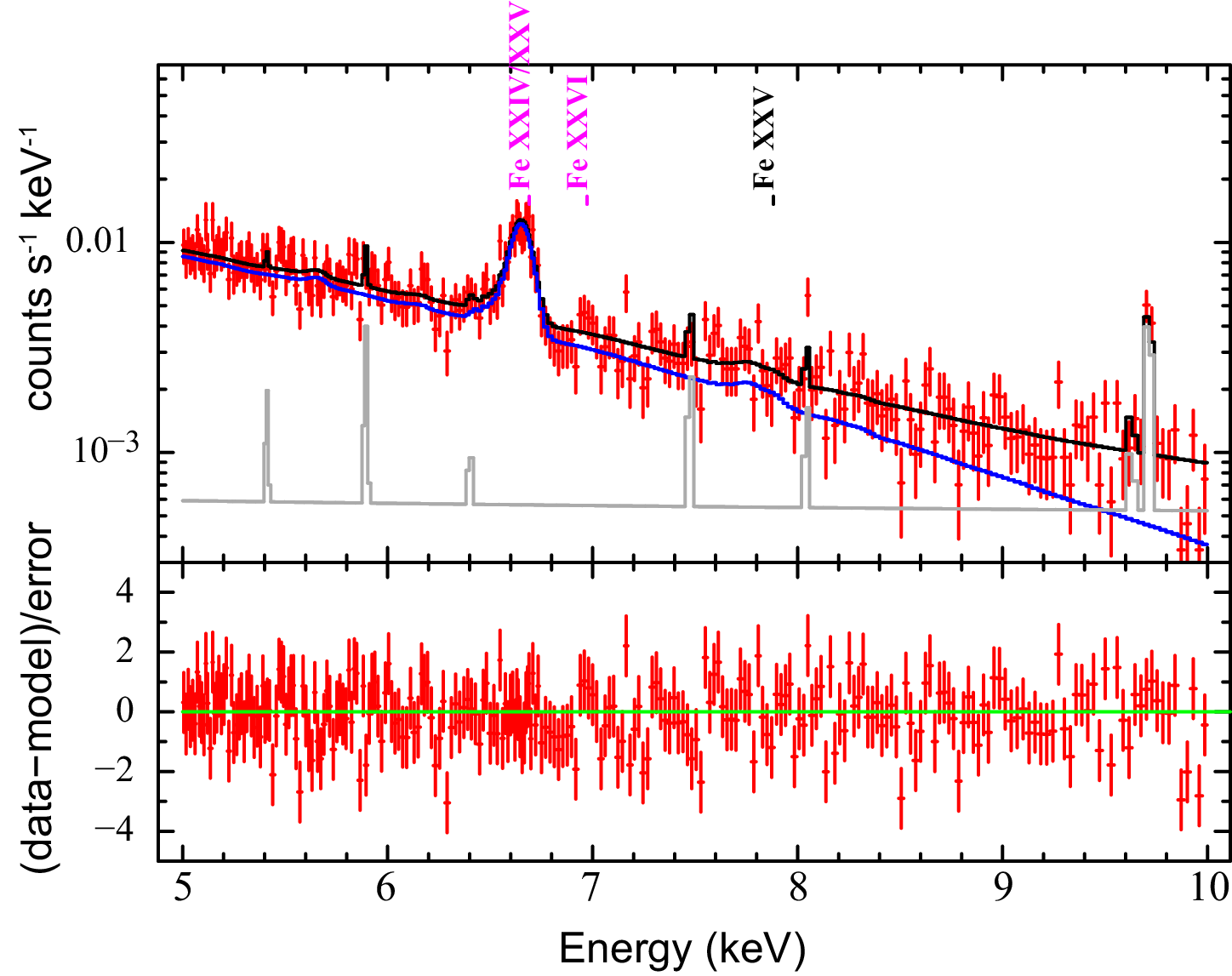}
    \end{center}
    \caption{The Fe-K line spectra are displayed using red crosses. The best-fit models, a power law with Gaussian (top) and the {\it bvpshock} model (bottom), are shown in blue line. The gray and the black lines indicate the NXB model and the combined result of the blue model and the NXB model. Spectral analyses for both models are conducted within an identical energy range of 5 -- 10 keV.
    {Alt text: Two line graphs showing both the data and the best-fit models with residuals.} 
     }
    \label{fig:sn1987a_FeK_spectra}
\end{figure}
%============== Figure 2 ====================

%=================================================================
\section{Analyses of the Spectral shape with XRISM}
\label{sec:data_analyses_spectrum}
%=================================================================
As demonstrated in Section \ref{sec:data_reduction_validation}, the X-ray spectra of SN~1987A were successfully acquired using XRISM Resolve. 
The objective of this section is to formulate the emission model of the high-resolution X-ray spectra obtained by XRISM and to derive plasma properties. 
This section is structured as follows.
First, we analyze the spectral features near the Fe-K line band with Resolve  (Section \ref{sec:data_analyses_spectrum:fe_band}).
Subsequently, an investigation of Resolve's spectra in lower energy bands is presented (Section \ref{sec:data_analyses_spectrum:resolve_3band}). 
Finally, a full spectral model with Resolve is established (Section \ref{sec:data_analyses_spectrum:resolve_all}).

%------------------------------------------------
\subsection{Fe-K line Profile with Resolve}
\label{sec:data_analyses_spectrum:fe_band}
%------------------------------------------------
In this section, we investigate the Resolve spectrum within the hard X-ray range of 5 -- 10 keV.
As illustrated in Figure \ref{fig:sn1987a_FeK_spectra}, Fe-K lines are clearly detected, as already reported by XMM-Newton since 2009 \citep{2010A&A...515A...5S,2012A&A...548L...3M}, NuSTAR since 2012 \citep{2021ApJ...916...76A} and Chandra since 2018 \citep{2024ApJ...966..147R}.
The X-ray spectrum is well described by a simple power law + Gaussian model yielding a C-stat of 1221.11 over 1132 dof, as shown in the left panel of Figure \ref{fig:sn1987a_FeK_spectra}.
The NXB model for Resolve, established in Section \ref{sec:data_reduction_validation:background}, was fixed during the fitting.
We obtained a power law photon index of $3.41 \pm 0.15$, and a line center energy at $6.644_{-0.008}^{+0.011}$ keV with a width of $58.2_{-6.2}^{+9.0}$ eV. 
The line equivalent width was $323_{-40}^{+60}$ eV, which corresponds to a line flux of $7.35_{-0.85}^{+1.00} \times 10^{-14}$ erg cm$^{-2}$ s$^{-1}$ or $6.92_{-0.80}^{+0.94} \times 10^{-6}$ photons cm$^{-2}$ s$^{-1}$.
The intensity of the Fe K line generally aligns with the long-term trend described by \citet{2021ApJ...916...41S}, \citet{2024ApJ...966..147R}, and \citet{2025ApJ...981...26S}. 
A comprehensive analysis of the temporal evolution of the plasmas will be provided in a future publication using Xtend and Resolve data.

Subsequently, we employed a more realistic plasma model, specifically a plane-parallel shock model (\cite{2001ApJ...548..820B}; implemented as the {\it bvpshock} model in XSPEC). 
This model represents X-ray emission from a hot plasma heated by a plane-parallel shock with a constant post-shock temperature and with a linear distribution in the ionization parameter $\tau$ across the emission measure.
In the fitting, the lower limit of the ionization parameter was fixed at 0 and the upper limit $\tau_{\rm u}$ was set to be free.
As described in Section \ref{sec:data_reduction_validation:background}, the photo absorption models by Galactic and LMC gas and the Doppler shift by the intrinsic motion of the object are multiplied with fixed values to the plasma model, and we used the abundance table for the LMC by \citet{1992ApJ...384..508R}.
Note that the following results for the redshift parameter ($z$) in the {\it bvpshock} model represent the redshift after accounting for intrinsic motion subtracted by {\it zashift} model.
The NXB model for Resolve is also frozen during this fitting.
The abundance values for lighter elements than Fe are set to those for 2019 November values in Table 4 of \citet{2021ApJ...916...41S}, but converted to the LMC equivalent values in our model. 
Here, the Ni abundance is tied to the Fe abundance in the fitting.
The X-ray spectrum is well described by the {\it bvpshock} model, as shown in the "pshock\_high(Sun+21)" column of Table \ref{tab:resolve_broad_fit_param_high}, with $kT = 3.17_{-0.25}^{+0.31}$ keV and $\tau_{\rm u} = (2.20_{-0.64}^{+0.75}) \times 10^{11}$ s cm$^{-3}$.
The Fe line complex is broadened by a Doppler velocity of $(16.1_{-5.1}^{+4.9})\times 10^2$ km s$^{-1}$, corresponding to a width of $32.2_{-10.2}^{+9.8}$ eV at 6 keV, and Doppler shifted by $z=(-14.9_{-24.4}^{+7.6})\times 10^{-4}$, which corresponds to a velocity of $(-4.5_{-7.3}^{+2.2})\times 10^2$ km s$^{-1}$ relative to the intrinsic motion of SN~1987A.
When utilizing alternative abundances for lighter elements from \citet{2010A&A...515A...5S} or those referenced in \citet{2024ApJ...966..147R}, the results remain in agreement within the statistical uncertainties, as demonstrated in the "pshock\_high(Sturm+10)" and "pshock\_high(Ravi+24)" columns presented in Table \ref{tab:resolve_broad_fit_param_high}.

%============== Table 1 ======================
\begin{table*}[bt]
    \tbl{Best-fit parameters for the X-ray spectrum of SN1987A with XRISM Resolve in 5-10 keV.}{%
    \begin{tabular}{lccc}
    \hline %------------------------------------------------------------------------------------------------------------------------------
    Model$^*$       & {\bf pshock\_high(Sun+21)$^\dagger$} & {\bf pshock\_high(Sturm+10)$^\ddagger$}   & {\bf pshock\_high(Ravi+24)$^\S$}      \\
%                      401h_sun21             401h_stm10                 401h_rav24                  
    \hline %------------------------------------------------------------------------------------------------------------------------------
    $kT$ (keV)      & $3.17_{-0.25}^{+0.31}$  & $3.17_{-0.25}^{+0.31}$  & $3.17_{-0.26}^{+0.30}$  \\    
    He              & 1.00 fix                & 1.00 fix                & 1.94 fix $^\S$\\    
    C               & 1.00 fix                & 1.00 fix                & 0.29 fix $^\S$\\
    N               & 5.78 fix $^\dagger$     & 7.62 fix  $^\ddagger$   & 4.51 fix $^\S$\\
    O               & 0.28 fix $^\dagger$     & 0.28 fix  $^\ddagger$   & 0.31 fix $^\S$\\
    Ne              & 0.73 fix $^\dagger$     & 0.72 fix  $^\ddagger$   & 0.71 fix $^\S$\\
    Mg              & 0.35 fix $^\dagger$     & 0.25 fix  $^\ddagger$   & 0.34 fix $^\S$\\
    Si              & 1.35 fix $^\dagger$     & 0.97 fix  $^\ddagger$   & 1.17 fix $^\S$\\
    S               & 1.52 fix $^\dagger$     & 1.11 fix  $^\ddagger$   & 1.05 fix $^\S$\\
    Ar              & 1.00 fix                & 1.00 fix                & 1.00 fix \\
    Ca              & 1.00 fix                & 1.00 fix                & 1.00 fix  \\
    Fe              & $0.77_{-0.15}^{+0.22}$  & $0.77_{-0.15}^{+0.22}$  & $0.95_{-0.18}^{+0.27}$ \\
   Ni  &            tied to Fe              & tied to Fe              &  tied to Fe\\
    $\tau_{\rm u}$ ($10^{11}$ s cm$^{-3}$) $^\|$    
                    & $2.20_{-0.64}^{+0.75}$  & $2.21_{-0.73}^{+0.72}$  & $2.22_{-0.64}^{+0.76}$ \\
    Doppler Shift $z$ ($10^{-4}$) $^\#$      
                    & $-14.9_{-24.4}^{+7.6}$  & $-14.9_{-24.4}^{+7.5}$ & $-14.7_{-24.1}^{+7.7}$ \\
    Broadening $v$ ($10^{2}$ km s$^{-1}$) $^{**}$
                    & $16.1_{-5.1}^{+4.9}$    & $16.2_{-5.1}^{+4.9}$    & $16.1_{-5.1}^{+5.0}$  \\
%    \hline %------------------------------------------------------------------------------------------------------------------------------
%    NXB norm.       & 0.96 fix                & 0.96 fix                & 0.96 fix               \\
    \hline %------------------------------------------------------------------------------------------------------------------------------
    C-stat          & 1211.84                 &   1211.84                & 1211.34                \\
    d.o.f           & 1131                    &   1131                   & 1131                    \\
    \hline %------------------------------------------------------------------------------------------------------------------------------
    \end{tabular}}
    \label{tab:resolve_broad_fit_param_high}
    \begin{tabnote}
    \footnotemark[$*$]        The Galactic and LMC absorption column densities were set at $6.0 \times 10^{20}$ and $2.2 \times 10^{21}$ atoms cm$^{-2}$ fixed, respectively, and the redshift determined by the recession velocity was fixed at 286.7 km s$^{-1}$ (see Section \ref{sec:data_reduction_validation:spectrum}). Abundances are relative to the LMC values given by \citet{1992ApJ...384..508R}. Fitting energy range is 5.0 -- 10 keV band. NXB normalization is fixed to be 0.96. The errors are  90\% confidence level.\\
    \footnotemark[$\dagger$]  Abundances are set to the values in Table 4, 2019 Nov in \citet{2021ApJ...916...41S}; N = 5.78, O = 0.28, Ne = 0.73, Mg = 0.35 of LMC abundance.\\
    \footnotemark[$\ddagger$] Abundances given in \citet{2010A&A...515A...5S} with XMM-Newton, relative to LMC value,\\
    \footnotemark[$\S$]       Abundances used in \citet{2024ApJ...966..147R}, relative to LMC value; He by \citet{2010ApJ...717.1140M}, C by \citet{1996ApJ...461..993F}, and O by \citet{2009ApJ...692.1190Z}.\\    
    \footnotemark[$\|$] Upper limit on ionization timescale ($\tau$) in units of $10^{11}$ s cm$^{-3}$.\\
    \footnotemark[$\#$] Doppler shift $z$ in units of $10^{-4}$. The intrinsic velocity of SN~1987A is subtracted.\\
    \footnotemark[$**$] Doppler broadening velocity in units of $10^2$ km s$^{-1}$.
    \end{tabnote}
\end{table*}
%============== Table 1 ======================

%============== Table 2 ======================
\begin{table*}[hbt]
    \tbl{Best-fit parameters for the X-ray spectrum of SN1987A with XRISM Resolve.}{%
    \begin{tabular}{lccccc}
    \hline %------------------------------------------------------------------------------------------------------------------------------
    Model$^*$       & {\bf pshock\_mid}       & {\bf pshock\_low}     &  {\bf pshock\_wide}    & {\bf pshock\_wide2} & {\bf pshock\_gsmooth} \\
%                     401m_sun21              401l_sun21               401w_sun21_nxbfix        401w_sun21_nxbfree    402w_sun21_nxbfree    
    Energy band (keV) & 3.0 -- 5.0            & 1.7 -- 3.0             &  1.7 -- 10.0          & 1.7 -- 10.0          & 1.7 -- 10.0 \\
    \hline %------------------------------------------------------------------------------------------------------------------------------
    $kT$ (keV)      & $2.88_{-0.31}^{+0.41}$  & $1.96_{-0.44}^{+0.65}$ & $2.85_{-0.08}^{+0.09}$& $2.84_{-0.08}^{+0.09}$ & $2.83_{-0.09}^{+0.09}$ \\
    Si              & 1.35 fix $^\dagger$     & $0.68_{-0.19}^{+0.24}$ & $1.23_{-0.26}^{+0.29}$& $1.21_{-0.26}^{+0.29}$ & $1.16_{-0.27}^{+0.28}$\\
    S               & $1.19_{-0.85}^{+1.01}$  & $0.82_{-0.16}^{+0.19}$ & $1.24_{-0.20}^{+0.22}$& $1.23_{-0.20}^{+0.22}$ & $1.21_{-0.09}^{+0.20}$ \\
    Ar              & $0.95_{-0.27}^{+0.35}$  & 1.00 fix               & $0.82_{-0.20}^{+0.22}$& $0.82_{-0.20}^{+0.22}$ & $0.82_{-0.20}^{+0.21}$ \\
    Ca              & $1.05_{-0.40}^{+0.46}$  & 1.00 fix               & $1.03_{-0.41}^{+0.45}$& $1.03_{-0.40}^{+0.41}$ & $1.06_{-0.43}^{+0.40}$ \\
    Fe             & 0.55 fix $^\dagger$     & 0.55 fix $^\dagger$    & $0.72_{-0.11}^{+0.15}$& $0.72_{-0.11}^{+0.15}$ & $0.74_{-0.15}^{+0.14}$ \\
    Ni  & tied to Fe             & tied to Fe              & tied to Fe & tied to Fe              & tied to Fe\\
   $\tau_{\rm u}$ ($10^{11}$ s cm$^{-3}$) $^\ddagger$    
                    & $4.15_{-2.63}^{+5.92}$  & $3.68_{-1.75}^{+5.66}$ & $2.63_{-0.50}^{+0.54}$& $2.62_{-0.48}^{+0.58}$ & $2.66_{-0.48}^{+0.57}$\\
    Doppler Shift $z$ ($10^{-4}$) $^\S$      
                    & $-8.04_{-9.29}^{+9.88}$ & $1.13_{-6.98}^{+7.44}$ & $-3.30_{-9.33}^{+6.87}$& $-3.46_{-9.17}^{+6.92}$& $-1.39_{-6.12}^{+5.22}$\\
    Broadening $v$ ($10^{2}$ km s$^{-1}$) $^\|$
                    & $9.8_{-2.4}^{+4.1}$     & $10.9_{-2.2}^{+3.0}$   & $14.5_{-2.7}^{+3.2}$  & $14.4_{-2.7}^{+3.2}$ & $17.0_{-3.7}^{+4.1}$ $^\#$ \\
    Index $\alpha$ $^{**}$ & --               & --                     & --                    & --                   & $1.27_{-0.17}^{+0.30}$  \\ 
    \hline %------------------------------------------------------------------------------------------------------------------------------
    NXB norm.       & 0.96 fix                & 0.96 fix               & 0.96 fix              & $0.99 \pm 0.03$       & $0.99_{-0.02}^{+0.03}$ \\
    \hline %------------------------------------------------------------------------------------------------------------------------------
    C-stat          &   554.28                & 367.79                 &  2182.91              &  2181.19+12467.96 $^{\dagger\dagger}$    & 2178.23+12467.82 $^{\dagger\dagger}$\\
    d.o.f           &   558                   & 372                    &  2074                 &  2074+16597 $^{\dagger\dagger}$          & 2073+16597 $^{\dagger\dagger}$\\
    \hline %------------------------------------------------------------------------------------------------------------------------------
    \end{tabular}}
    \label{tab:resolve_broad_fit_param_sun21}
    \begin{tabnote}
    \footnotemark[$*$]        The Galactic and LMC absorption column densities were set at $6.0 \times 10^{20}$ and $2.2 \times 10^{21}$ atoms cm$^{-2}$ fixed, respectively, and the redshift determined by the recession velocity was fixed at 286.7 km s$^{-1}$ (see Section \ref{sec:data_reduction_validation:spectrum}). Abundances are relative to the LMC values given by \citet{1992ApJ...384..508R}. Abundances are set to the values in Table 4, 2019 Nov in \citet{2021ApJ...916...41S}; N = 5.78, O = 0.28, Ne = 0.73, Mg = 0.35 of LMC abundance. The errors are 90\% confidence level.\\
    \footnotemark[$\dagger$]  Abundances of 2019 Nov in Table 4 of \citet{2021ApJ...916...41S}.\\
    \footnotemark[$\ddagger$] Upper limit on ionization timescale ($\tau_{\rm u}$) in units of $10^{11}$ s cm$^{-3}$.\\
    \footnotemark[$\S$]       Doppler shift $z$ in units of $10^{-4}$. The intrinsic velocity of SN~1987A is subtracted.\\
    \footnotemark[$\|$]       Doppler broadening velocity in units of $10^2$ km s$^{-1}$.\\
    \footnotemark[$\#$]       Doppler broadening velocity at 6.0 keV. \\
    \footnotemark[$**$]       Power of energy for Doppler broadening variation\\
    \footnotemark[$\dagger\dagger$]  The initial and secondary values denote the source and NXB data, respectively.   \\
    \end{tabnote}
\end{table*}
%============== Table 2 ======================

Slight indications of Fe XXVI lines (located at 6.952 keV and 6.973 keV in the rest frame) appear in the residuals shown in the lower panel of Fig. \ref{fig:sn1987a_FeK_spectra}. 
By adding a model with two Gaussian functions where the line center separation and the intensity ratio were fixed at their theoretical values (21 eV and $\times 1.98$, respectively) on {\it pshock} model above, the fit slightly improved with a C-stat of 1209.13 for 1128 d.o.f., although it is not statistically significant (for reference, F-statistic value is 0.8 with a probability of 0.4). 
The 90\% upper limit for the equivalent width of the sum of the two lines was determined as $<17.0$ eV. 
The line center energies are $6.967_{-0.033}^{+0.202}$ keV and $6.988$ keV with the same error, indicating a Doppler shift of $z=-22.0_{-145.4}^{+46.9}\times10^{-4}$ or $6.6^{+14.1}_{-43.6}\times10^2$ km s$^{-1}$. 
The upper limit for the line width is $<70.3$ eV, corresponding to a Doppler broadening velocity of $<3.03\times 10^{3}$ km s$^{-1}$. 
Both the Doppler broadening and shift align with those of the {\it bvpshock} plasma outlined in Table \ref{tab:resolve_broad_fit_param_high}.

%============== Figure 3 ====================
\begin{figure}[hbt]
    \begin{center}
    \includegraphics[width=0.48 \textwidth]{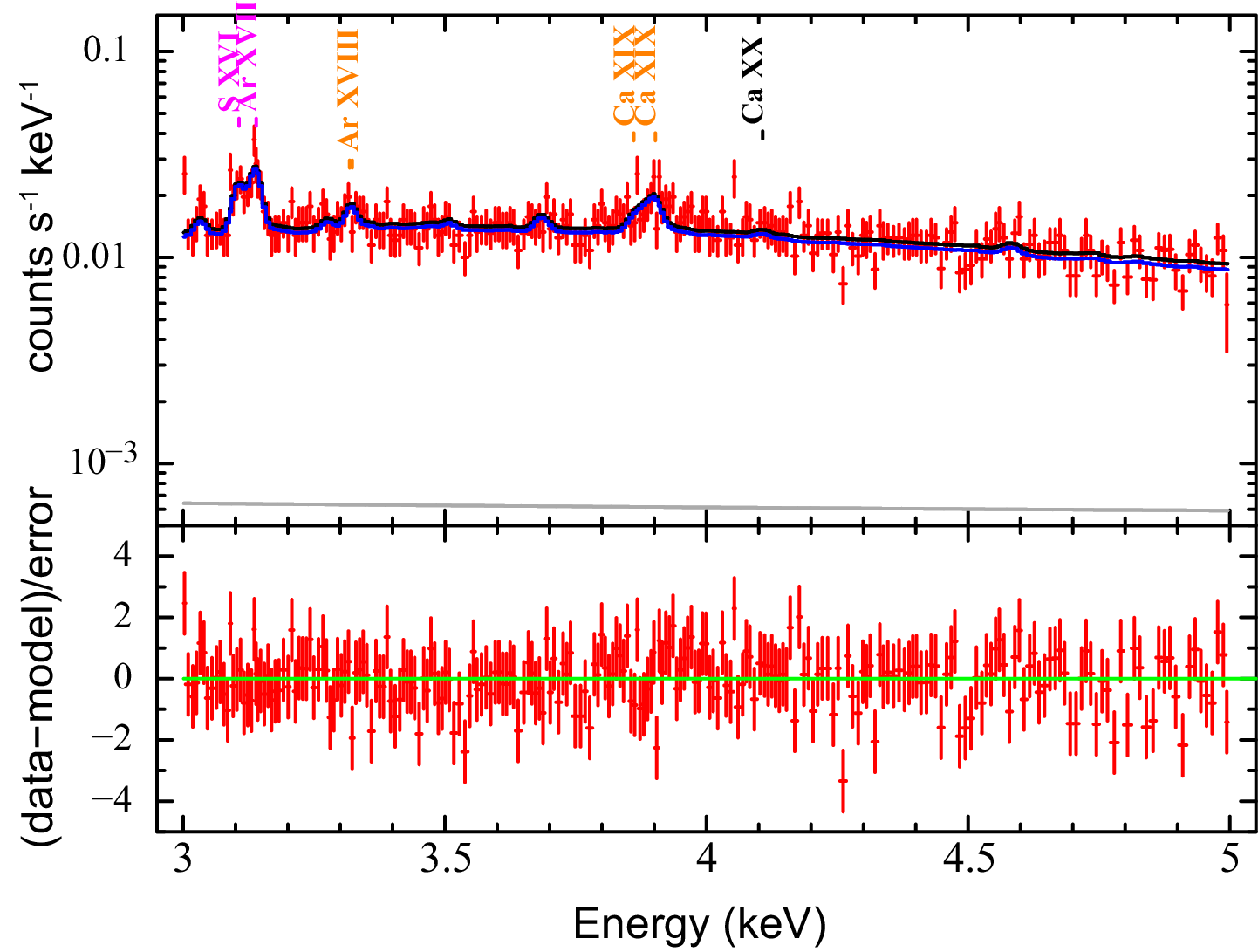}
    \includegraphics[width=0.48 \textwidth]{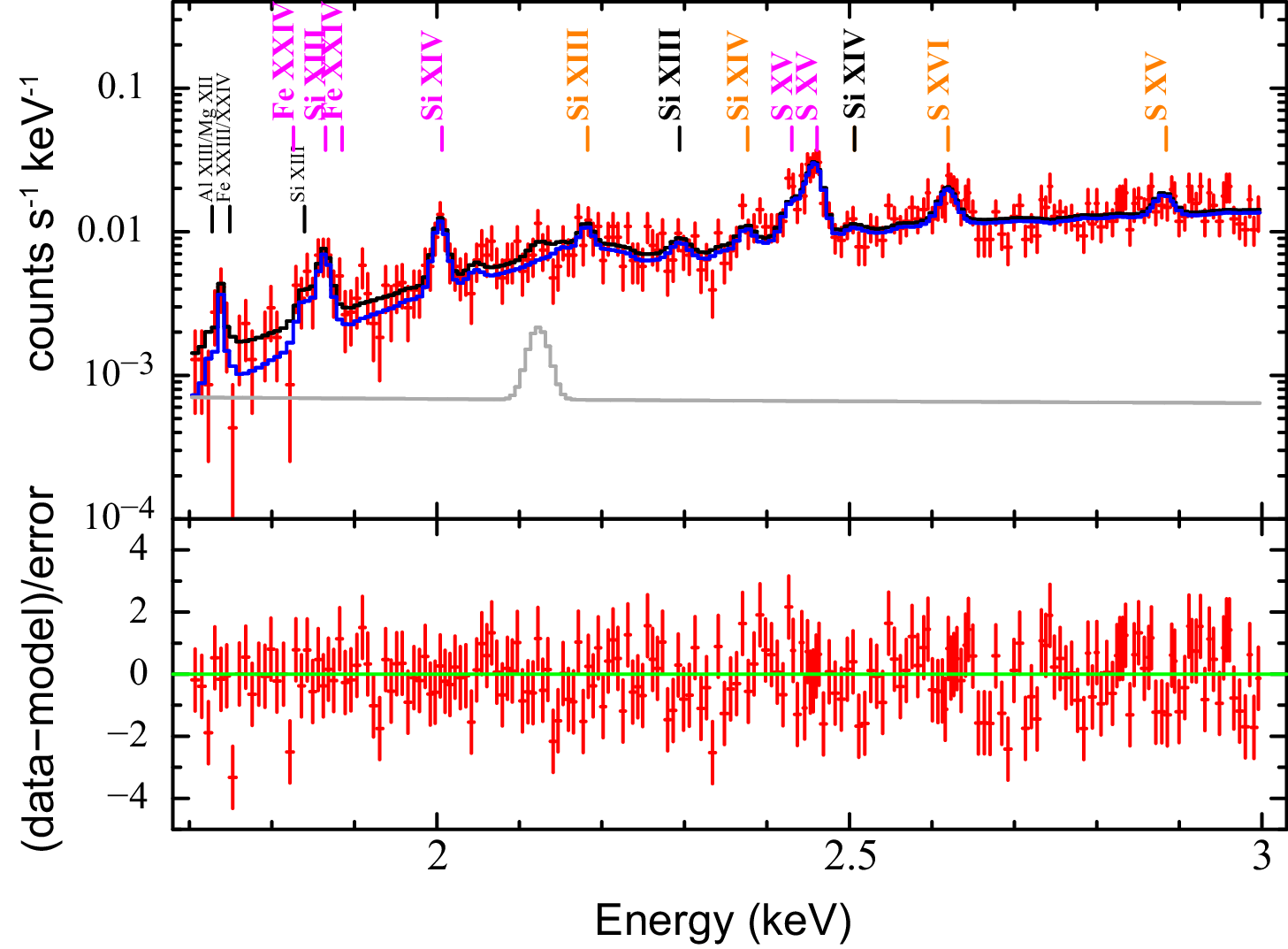}
    \end{center}
    \caption{The X-ray spectra of SN1987A with Resolve in the 1.7 -- 3 keV, 3 -- 5 keV bands are shown in red crosses in the top panels of the top, and bottom plots, respectively. 
    The best-fit models of "pshock\_mid" and "pshock\_low" in Table \ref{tab:resolve_broad_fit_param_sun21} are also shown in blue lines in the top panels, respectively, and the NXB model and the sum of the best-fit model and NXB model are shown in gray and black, respectively.
    The bottom panel in each plot represents the delta $\chi^2$.
    {Alt text: Two line graphs showing both the data and the best-fit models with residuals.} 
     }
    \label{fig:sn1987a_narrow_spectra}
\end{figure}
%============== Figure 3 ====================

%------------------------------------------------
\subsection{Resolve spectral feature in the lower energy bands}
\label{sec:data_analyses_spectrum:resolve_3band}
%------------------------------------------------
As seen in Section \ref{sec:data_analyses_spectrum:fe_band}, the Resolve spectrum in the 5.0 -- 10 keV band is well reproduced by the {\it pshock} model family.  We next examined the spectral shape in the lower energy band.
We applied the same plasma model as in Section \ref{sec:data_analyses_spectrum:fe_band}, {\it bvpshock} model multiplied and/or convolved by {\it wabs*phabs*zashift} with fixed values in Section \ref{sec:data_reduction_validation:spectrum} on the Resolve spectra in the 3.0 -- 5.0 keV and the 1.7 -- 3.0 keV bands.
In the fitting, the NXB model for Resolve was frozen with the constant factor of 0.96 (Section \ref{sec:data_reduction_validation:background}).
The results for the 3.0 -- 5.0 keV and the 1.7 -- 3.0 keV bands are shown in the top and bottom panels of Figure \ref{fig:sn1987a_narrow_spectra} and "pshock\_mid" and "pshock\_low" models in Table \ref{tab:resolve_broad_fit_param_sun21}, respectively.
The abundances of elements whose line energies are out of the fit energy range were fixed at the value of 2019 November in Table 4 of \citet{2021ApJ...916...41S}, as we did in Section \ref{sec:data_analyses_spectrum:fe_band}.
As a result, we found that all three narrow-band spectra are well represented by a single {\it bvpshock} model.
The $kT$ and $\tau_{\rm u}$ obtained from "pshock\_mid" and "pshock\_low" models are consistent with those by "pshock\_high" model in Section \ref{sec:data_analyses_spectrum:fe_band} within the statistical errors.
This conclusion does not change within the statistical errors even when applying alternative abundance values from \citet{2010A&A...515A...5S} or those used in \citet{2024ApJ...966..147R}.  % No Table.

%============== Figure 4 ====================
\begin{figure}[hbt]
    \begin{center}
    \includegraphics[width=0.48 \textwidth]{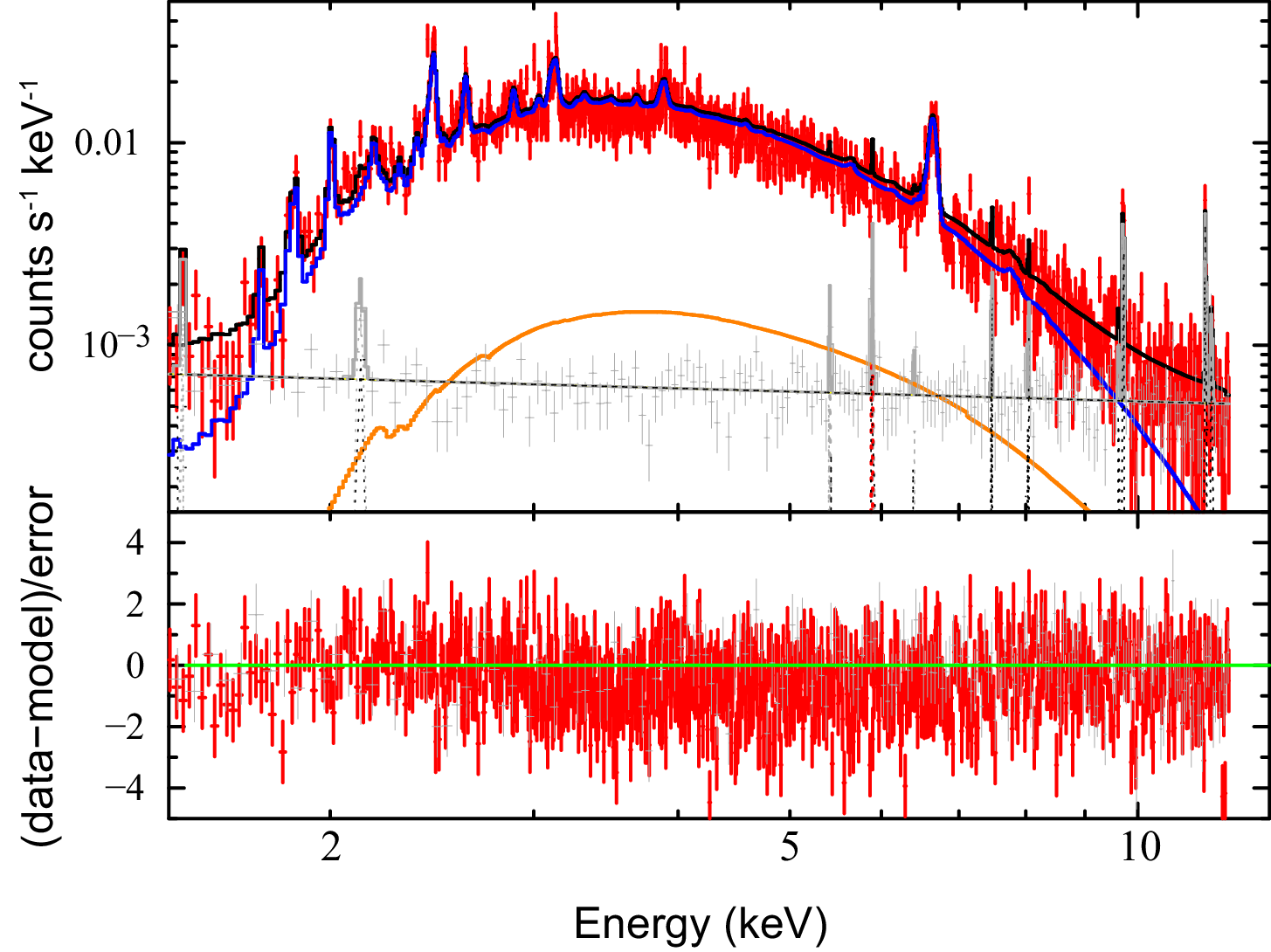} 
    \end{center}
    \caption{(top) X-ray spectra of SN1987A and the non-X-ray background with XRISM Resolve are shown in red and gray crosses, respectively.
    The blue line illustrates the best-fit model for "pshock\_wide2" as detailed in Table \ref{tab:resolve_broad_fit_param_sun21}.  
    The gray and black lines indicate the best-fit models for the NXB alone and the combined model of the NXB and "pshock\_wide2", respectively.
    The orange line represents the pulsar wind nebula model suggested by \citet{2022ApJ...931..132G} (See the text in Section \ref{sec:discussion:pwn} later).
    (bottom) The residuals for the SN~1987A data and the NXB spectra are displayed in red and gray, respectively.
    {Alt text: One line graph illustrating both the data and the best-fit models with residuals.} 
    }
    \label{fig:sn1987a_broad_spectra}
\end{figure}
%============== Figure 4 ====================

%============== Figure 5 ====================
\begin{figure}[htb]
    \begin{center}
        \includegraphics[width=0.45 \textwidth]{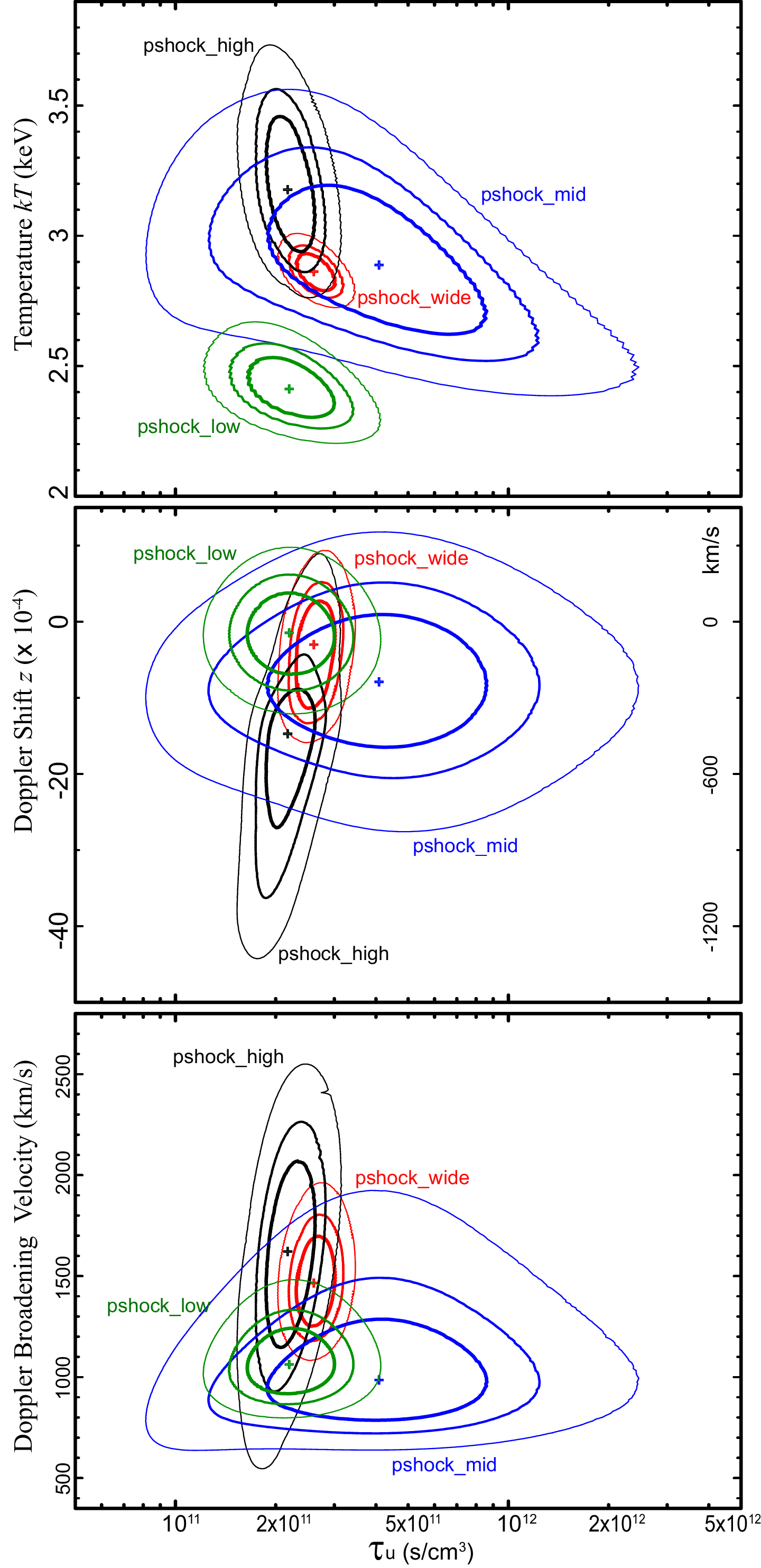}
    \end{center}
    \caption{In the top, middle, and bottom panels, the confidence contours at 68\% (1$\sigma$), 90\%, and 99\% levels between $\tau_{\rm u}$ and $kT$, $z$, and Doppler velocity, respectively, for "pshock\_wide", "pshock\_high", "pshock\_mid", and "pshock\_low" from Table \ref{tab:resolve_broad_fit_param_sun21} are displayed in red, green, blue, and black. The best-fit parameters are indicated by plus signs. These plots are derived from fitting the spectra with parameters other than $kT$, $z$, and velocity held constant.  
    {Alt text: Three line graphs showing confidence contours across four parameters, overlapping for four different model fittings.}
     }
    \label{fig:sn1987a_3band_cont}
\end{figure}
%============== Figure 5 ====================

%============== Table 3 ======================
\begin{table*}[hbt]
    \tbl{Same table as Table \ref{tab:resolve_broad_fit_param_sun21} but with different abundance parameters.}{%
    \begin{tabular}{lcccccc}
    \hline %------------------------------------------------------------------------------------------------------------------------------
    Model$^*$       & {\bf pshock\_wide}      & {\bf pshock\_wide2}     & {\bf pshock\_gsmooth}   &  {\bf pshock\_wide}   & {\bf pshock\_wide2}    & {\bf pshock\_gsmooth} \\
%%                    401w_stm10_nxbfix         401w_stm10_nxbfree      402w_stm10_nxbfree        401w_rav24_nxbfix      401w_rav24_nxbfree        402w_rav24_nxbfree    
    \hline %------------------------------------------------------------------------------------------------------------------------------
    $kT$ (keV)      & $2.86_{-0.08}^{+0.09}$  & $2.85_{-0.08}^{+0.09}$  & $2.83_{-0.07}^{+0.06}$ & $2.85_{-0.08}^{+0.09}$& $2.84_{-0.08}^{+0.09}$  & $2.82_{-0.08}^{+0.09}$ \\
    He              & 1.00 fix                & 1.00 fix                & 1.00 fix               & 1.94 fix $^\ddagger$   & 1.94 fix $^\ddagger$   & 1.94 fix $^\ddagger$ \\
    C               & 1.00 fix                & 1.00 fix                & 1.00 fix               & 0.29 fix $^\ddagger$   & 0.29 fix $^\ddagger$   & 0.29 fix $^\ddagger$ \\
    N               & 7.62 fix $^\dagger$     & 7.62 fix $^\dagger$     & 7.62 fix $^\dagger$    & 4.51 fix $^\ddagger$   & 4.51 fix $^\ddagger$   & 4.51 fix $^\ddagger$ \\
    O               & 0.28 fix $^\dagger$     & 0.28 fix $^\dagger$     & 0.28 fix $^\dagger$    & 0.31 fix $^\ddagger$   & 0.31 fix $^\ddagger$   & 0.31 fix $^\ddagger$\\
    Ne              & 0.72 fix $^\dagger$     & 0.72 fix $^\dagger$     & 0.72 fix $^\dagger$    & 0.71 fix $^\ddagger$   & 0.71 fix $^\ddagger$   & 0.71 fix $^\ddagger$\\
    Mg              & 0.25 fix $^\dagger$     & 0.25 fix $^\dagger$     & 0.25 fix $^\dagger$    & 0.34 fix $^\ddagger$   & 0.34 fix $^\ddagger$   & 0.34 fix $^\ddagger$ \\
    Si              & $1.23_{-0.26}^{+0.29}$  & $1.21_{-0.26}^{+0.29}$  & $1.16_{-0.21}^{+0.23}$ & $1.53_{-0.32}^{+0.36}$ & $1.51_{-0.32}^{+0.36}$ & $0.88_{-0.20}^{+0.21}$\\
    S               & $1.24_{-0.20}^{+0.22}$  & $1.23_{-0.20}^{+0.22}$  & $1.21_{-0.15}^{+0.16}$ & $1.54_{-0.25}^{+0.27}$ & $1.53_{-0.25}^{+0.27}$ & $0.92_{-0.13}^{+0.15}$ \\
    Ar              & $0.82_{-0.20}^{+0.22}$  & $0.82_{-0.20}^{+0.22}$  & $0.82_{-0.20}^{+0.21}$ & $1.02_{-0.25}^{+0.28}$ & $1.02_{-0.25}^{+0.28}$ & $0.62_{-0.15}^{+0.16}$ \\
    Ca              & $1.03_{-0.41}^{+0.45}$  & $1.03_{-0.41}^{+0.45}$  & $1.07_{-0.41}^{+0.43}$ & $1.28_{-0.51}^{+0.55}$ & $1.27_{-0.51}^{+0.56}$ & $0.81_{-0.31}^{+0.32}$ \\
    Fe              & $0.72_{-0.11}^{+0.15}$  & $0.72_{-0.11}^{+0.16}$  & $0.76_{-0.09}^{+0.09}$ & $0.89_{-0.14}^{+0.19}$ & $0.90_{-0.14}^{+0.20}$ & $0.57_{-0.09}^{+0.09}$ \\
    Ni  & tied to Fe             & tied to Fe              & tied to Fe & tied to Fe              & tied to Fe              & tied to Fe \\
    $\tau_{\rm u}$ ($10^{11}$ s cm$^{-3}$) $^\S$    
                    & $2.64_{-0.51}^{+0.54}$  & $2.62_{-0.49}^{+0.58}$  & $2.65_{-0.42}^{+0.42}$ & $2.62_{-0.48}^{+0.58}$ & $2.65_{-0.51}^{+0.55}$ & $2.64_{-0.37}^{+0.53}$\\
    Doppler Shift $z$ ($10^{-4}$) $^\|$      
                    & $-3.34_{-9.33}^{+6.91}$ & $-3.51_{-9.18}^{+6.98}$ & $-3.33_{-2.56}^{+7.80}$& $-3.50_{-9.07}^{+6.92}$& $-3.62_{-9.02}^{+6.94}$& $-3.34_{-4.39}^{+7.00}$\\
    Broadening $v$ ($10^{2}$ km s$^{-1}$) $^\#$
                    & $14.6_{-2.7}^{+3.2}$    & $14.5_{-2.7}^{+3.2}$    & $18.7_{-2.2}^{+2.2}$ $^{**}$ & $14.5_{-2.7}^{+3.2}$& $14.4_{-2.7}^{+3.2}$  & $17.0_{-3.6}^{+3.0}$ $^{**}$ \\
    Index $\alpha$ $^{\dagger\dagger}$ & --   & --                      & $1.35_{-0.29}^{+0.19}$    & --                    & --                   & $1.26_{-0.32}^{+0.31}$  \\ 
    \hline %------------------------------------------------------------------------------------------------------------------------------
    NXB norm.       & 0.96 fix                & $0.99 \pm 0.03$         & $0.99 \pm 0.02$        & 0.96 fix               & $0.99 \pm 0.03$         & $0.99_{-0.03}^{+0.01}$ \\
    \hline %------------------------------------------------------------------------------------------------------------------------------
    C-stat          & 2183.04                 &   2181.30+12467.96$^{\dagger\dagger}$      & 2178.75+12467.82$^{\dagger\dagger}$       
                                                                                            &  2182.31              &  2180.57+12467.97$^{\dagger\dagger}$       & 2181.10+12467.80$^{\dagger\dagger}$\\
    d.o.f           & 2074                    &   2074+16597$^{\dagger\dagger}$            & 2073+16597$^{\dagger\dagger}$             
                                                                                            &  2074                 &  2074+16597$^{\dagger\dagger}$             & 2073+16597$^{\dagger\dagger}$ \\
    \hline %------------------------------------------------------------------------------------------------------------------------------
    \end{tabular}}
    \label{tab:resolve_broad_fit_param_stm10_rav24}
    \begin{tabnote}
    \footnotemark[$*$]        The Galactic and LMC absorption column densities were set at $6.0 \times 10^{20}$ and $2.2 \times 10^{21}$ atoms cm$^{-2}$ fixed, respectively, and the redshift determined by the recession velocity was fixed at 286.7 km s$^{-1}$ (see Section \ref{sec:data_reduction_validation:spectrum}). Abundances are relative to the LMC values given by \citet{1992ApJ...384..508R}. Fitting energy range is 1.5 -- 10 keV band. The errors are 90\% confidence level.\\
    \footnotemark[$\dagger$]  Abundances given in \citet{2010A&A...515A...5S} with XMM-Newton, relative to LMC value,\\
    \footnotemark[$\ddagger$] Abundances used in \citet{2024ApJ...966..147R}, relative to LMC value; He by \citet{2010ApJ...717.1140M}, C by \citet{1996ApJ...461..993F}, and O by \citet{2009ApJ...692.1190Z}.\\
    \footnotemark[$\S$]       Upper limit on ionization timescale ($\tau_{\rm u}$) in units of $10^{11}$ s cm$^{-3}$.\\
    \footnotemark[$\|$]       Doppler shift $z$ in units of $10^{-4}$. The intrinsic velocity of SN~1987A is subtracted.\\
    \footnotemark[$\#$]       Doppler broadening velocity in units of $10^2$ km s$^{-1}$.\\
    \footnotemark[$**$]    Doppler broadening at 6.0 keV. \\
    \footnotemark[$\dagger\dagger$]  Power of energy for Doppler broadening variation.\\
    \footnotemark[$\ddagger\ddagger$]  The initial and secondary values denote the source and NXB data, respectively.   \\
    \end{tabnote}
\end{table*}
%============== Table 3 ======================

%============== Table 4 ======================
\begin{table*}[hbt]
    \tbl{Best-fit parameters for the X-ray lines in Resolve spectra.}{%
    \begin{tabular}{lcccccc}
    \hline %-----------------------------------------------------------------------------------------------------------------------------------------------------
    Target          & {\bf Fe XXV}          & {\bf Ar XVII/S XVI}       & {\bf S XVI}           &  {\bf S XV}             & {\bf Si XIV}           & {\bf Si XIII} \\
    %% Note         60_74_fe_xxv            30_33_s_xvi                 25_27_s_xvi             24_25_s_xv                19_20_si_xiv             18_19_si_xiii
    Energy band (keV)   & 6.0 -- 7.4        & 3.00 -- 3.25              & 2.50 -- 2.74          & 2.40 -- 2.52            & 1.94 -- 2.08           & 1.80 -- 1.91 \\
    \hline %------------------------------------------------------------------------------------------------------------------------------------------------------
    \multicolumn{7}{l}{\bf Abundances Fixed$^*$}\\
    Doppler Shift $z$ ($10^{-4}$)$^\S$
                   & $-14.1_{-8.9}^{+6.0}$  & $-6.8_{-9.1}^{+9.1}$      & $4.3_{-28.8}^{+40.0}$ & $7.8_{-8.5}^{+10.0}$  & $0.5_{-19.9}^{+20.4}$  & $-4.1_{-27.0}^{+21.2}$\\
    Broadening $v$ ($10^{2}$ km s$^{-1}$)$^\|$
                   & $16.5_{-4.7}^{+4.6}$   & $8.0_{-1.9}^{+2.7}$       & $21.4_{-6.8}^{+9.9}$  & $10.9_{-2.3}^{+3.4}$    & $15.4_{-4.6}^{+7.8}$   & $12.2_{-5.3}^{+10.4}$\\
    C-stat         &  333.10                &  53.55                    & 63.99                 & 37.74                   &   34.99          & 32.23\\
    d.o.f          &  344                   &  68                       & 64                    & 30                      &   40                & 28\\
    \hline %------------------------------------------------------------------------------------------------------------------------------------------------------
    \multicolumn{7}{l}{\bf Abundances Free$^\dagger$}\\
    $z$ ($10^{-4}$)$^\S$ 
                   & $-13.9_{-8.5}^{+6.0}$  & $-6.6_{-9.3}^{+9.4}$      & $7.5_{-32.2}^{+50.6}$  & $7.8_{-8.6}^{+10.5}$   & $-2.2_{-18.5}^{+20.3}$ & $-1.7_{-28.8}^{+19.8}$\\
    Velocity ($10^{2}$ km s$^{-1}$)$^\|$  
                   & $17.2_{-5.0}^{+4.9}$   & $7.5_{-2.2}^{+2.9}$       & $22.9_{-10.2}^{+14.7}$ & $10.8_{-2.7}^{+4.2}$   & $11.1_{-5.9}^{+6.6}$   & $10.6_{-5.2}^{+15.9}$ \\
    Abundance      & $0.76_{-0.12}^{+0.13}$ & $0.82_{-0.75}^{+0.89}$    & $1.37_{-0.62}^{+1.15}$ & $1.22_{-0.34}^{+0.47}$ & $0.60_{-0.28}^{+0.38}$ & $1.00_{-0.57}^{+2.03}$\\
    Abundance      & --                     & $0.75_{-0.22}^{+0.25}$ $^\ddagger$       & --           & --                    & -- & --\\
    C-stat         &  332.68                &  52.65                    & 63.87                  & 37.74                 &   29.19          & 32.10\\
    d.o.f          &  343                   &  66                       & 63                     & 29                    &   39                & 27\\
    \hline %------------------------------------------------------------------------------------------------------------------------------
    \end{tabular}}
    \label{tab:resolve_line_fit_param}
    \begin{tabnote}
    \footnotemark[$*$]  The Galactic and LMC absorption column densities were set at $6.0 \times 10^{20}$ and $2.2 \times 10^{21}$ atoms cm$^{-2}$ fixed, respectively, and the redshift determined by the recession velocity was fixed at 286.7 km s$^{-1}$ (see Section \ref{sec:data_reduction_validation:spectrum}). All abundance parameters are frozen at "pshock\_wide2" model in Table \ref{tab:resolve_broad_fit_param_sun21}.\\
    \footnotemark[$\dagger$] Abundance parameter for the interest nuclei is set to free, but others are frozen at "pshock\_wide2" model in Table \ref{tab:resolve_broad_fit_param_sun21}.\\
    \footnotemark[$\ddagger$] Abundance for S for the "Ar XVII/S XVI" fitting.\\
    \footnotemark[$\S$] $z$ is the Doppler shift in units of $10^{-4}$. 
    \footnotemark[$\|$] Doppler broadening velocity in units of $10^2$ km s$^{-1}$.\\
    \end{tabnote}
\end{table*}
%============== Table 4 ======================
%------------------------------------------------
\subsection{Resolve spectrum in the 1.7 -- 10 keV band}
\label{sec:data_analyses_spectrum:resolve_all}
%------------------------------------------------
Given that the Resolve spectra within the energy ranges of 1.7 -- 3 keV, 3 -- 5 keV, and 5 -- 10 keV are well represented by the {\it bvpshock} model, as detailed in Sections \ref{sec:data_analyses_spectrum:fe_band} and \ref{sec:data_analyses_spectrum:resolve_3band}, we applied the same plasma model (i.e., {\it wabs*phabs*zashift*bvpshock} in XSPEC) to the entire Resolve spectra across the 1.7 -- 10 keV range. 
The resultant spectra are well represented using this model, as illustrated in Figure \ref{fig:sn1987a_broad_spectra} and Table \ref{tab:resolve_broad_fit_param_sun21}. 
It is noteworthy that the statistical uncertainties in the Resolve NXB spectrum do not affect this outcome, as highlighted by the "pshock\_wide" and "pshock\_wide2" models in Table \ref{tab:resolve_broad_fit_param_sun21}, corresponding to the NXB free and fixed fits, respectively. 
For these fits, the abundance values of elements lighter than Fe were taken from Table 5 of \citet{2021ApJ...916...41S}; the results remain consistent even when alternative abundances from \citet{2021ApJ...916...41S} or \citet{2024ApJ...966..147R} are used, as shown in Table \ref{tab:resolve_broad_fit_param_stm10_rav24}.

In order to verify the consistency of the plasma parameters across the "pshock\_low", "pshock\_mid", "pshock\_high", and "pshock\_wide" models in Tables \ref{tab:resolve_broad_fit_param_high}, \ref{tab:resolve_broad_fit_param_sun21}, and \ref{tab:resolve_broad_fit_param_stm10_rav24}, we have plotted the confidence contours of $kT$, $z$, and velocity against $\tau_{\rm u}$ in Figure \ref{fig:sn1987a_3band_cont}.
The results demonstrate statistical consistency among these models within the 99\% confidence level, indicating that the single plasma component is dominant in the Resolve spectrum, although the spectrum of the lowest energy band typically favors a lower $kT$ value than others with less significance ( Figure \ref{fig:sn1987a_3band_cont} top). 
The Doppler shift $z$ (Figure \ref{fig:sn1987a_3band_cont} middle) is statistically consistent across all four results, exhibiting a clear positive correlation with $\tau_{\rm u}$ in both the high and wide-band fittings, a behavior characteristic of non-equilibrium coronal plasma.
The absolute value of $z$ from the intrinsic velocity is consistent with 0 km s$^{-1}$ within errors.

Examining the 90\% confidence level contours in Figure \ref{fig:sn1987a_3band_cont} (bottom), larger Doppler broadening on atomic lines may be required for the Fe-K (highest) energy band, compared with those in other bands.
The {\it bvpshock} model utilized in the previous analysis assumes a Doppler broadening energy power $\alpha = 1$ (i.e., constant velocity over all the atomic lines), expressed by $\Delta E \propto E^\alpha$ for the broadening width $\Delta E$ at line energy $E$. 
Therefore, in order to study the energy dependence of Doppler broadening, we applied the {\it gsmooth*vpshock} model in XSPEC, and obtained the power of energy of $\alpha = 1.25_{-0.37}^{+0.32}$, which is smaller than the previous X-ray results ($\alpha\sim2$) by \citet{2021ApJ...916...41S}.
This result remains consistent even when applying the elemental abundances lighter than Fe from \citet{2021ApJ...916...41S} or \citet{2024ApJ...966..147R}, as shown in Table \ref{tab:resolve_broad_fit_param_stm10_rav24}. 
This trend will be explored further in Section \ref{sec:data_analyses_line}.

%=================================================================
\section{Atomic Line Diagnostics using Resolve spectrum}
\label{sec:data_analyses_line}
%=================================================================
In order to measure the Doppler shift and broadening of atomic lines in the Resolve spectrum, we performed narrow band fitting around the line center energies using the {\it bvpshock} model whose $kT$ and $\tau_{\rm u}$ were frozen at "pshock\_wide2" model values in Table \ref{tab:resolve_broad_fit_param_sun21}.
Although the correlation between $z$ and $\tau_{\rm u}$ seen in Section \ref{sec:data_analyses_spectrum:resolve_3band} may affect the measurement of $z$, we fixed $\tau_{\rm u}$ here because it is consistent among three band fits in Section \ref{sec:data_analyses_spectrum}.
We tested all the atomic lines indicated in magenta and orange color labels in Figure \ref{fig:sn1987a_FeK_spectra} bottom and Figure \ref{fig:sn1987a_narrow_spectra} top and bottom panels. 
Among them, we obtained statistically significant parameters for six lines from Si, S, Ar, and Fe, which are shown in magenta labels in these figures.

%============== Figure 6 ====================
\begin{figure*}[htb]
    \begin{center}
    \includegraphics[width=0.48 \textwidth]{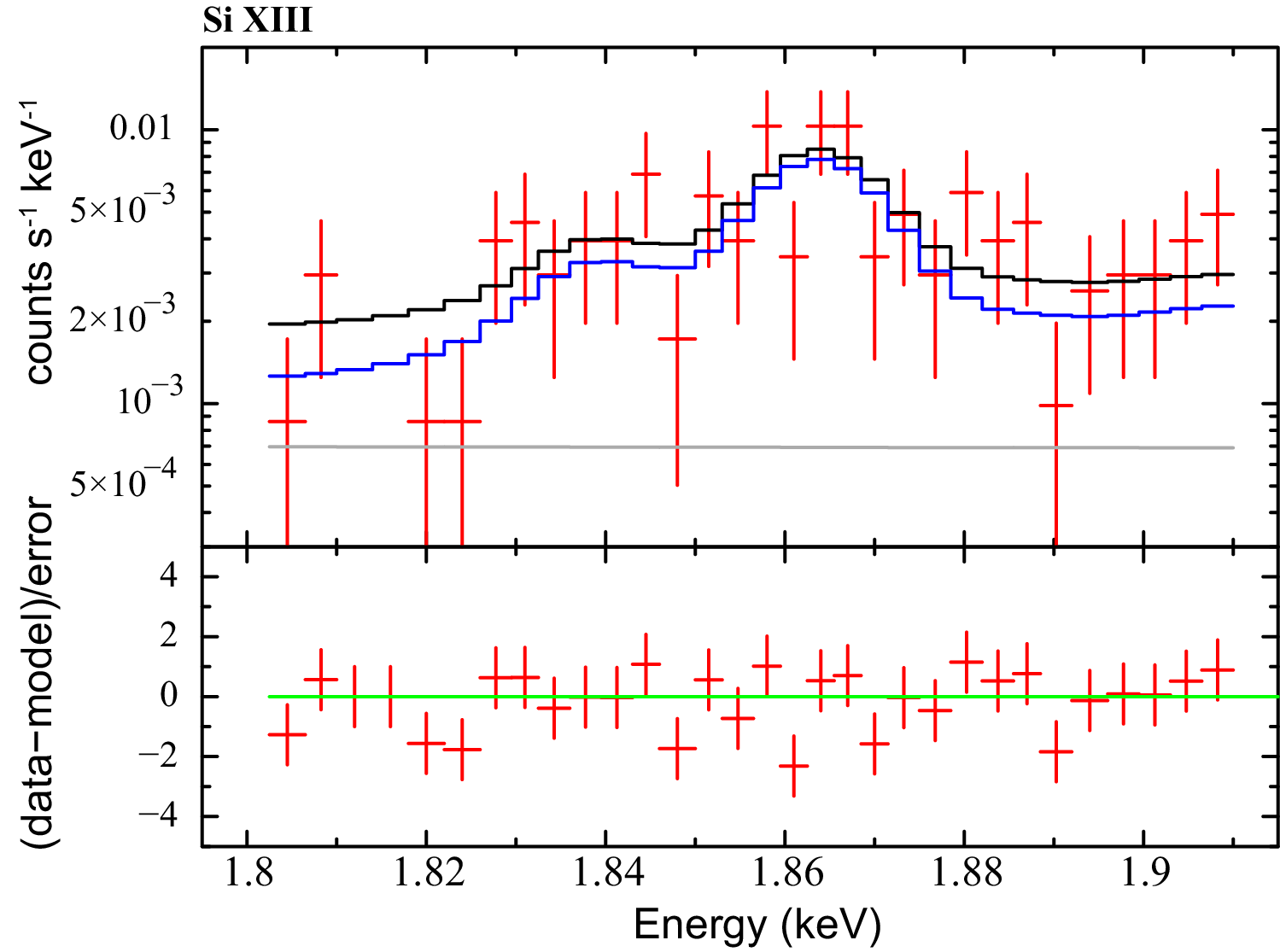}
    \includegraphics[width=0.48 \textwidth]{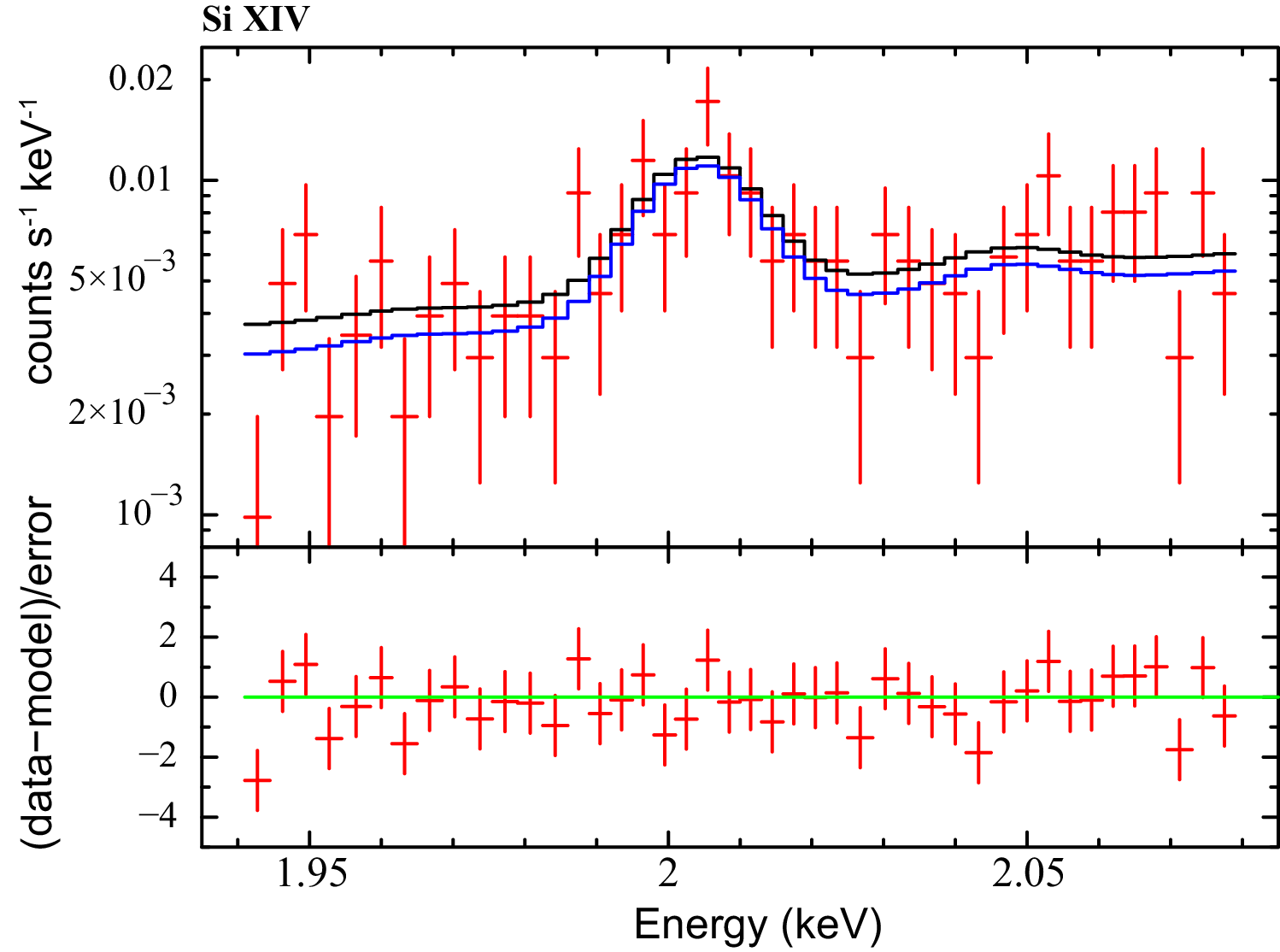}\\
    \includegraphics[width=0.48 \textwidth]{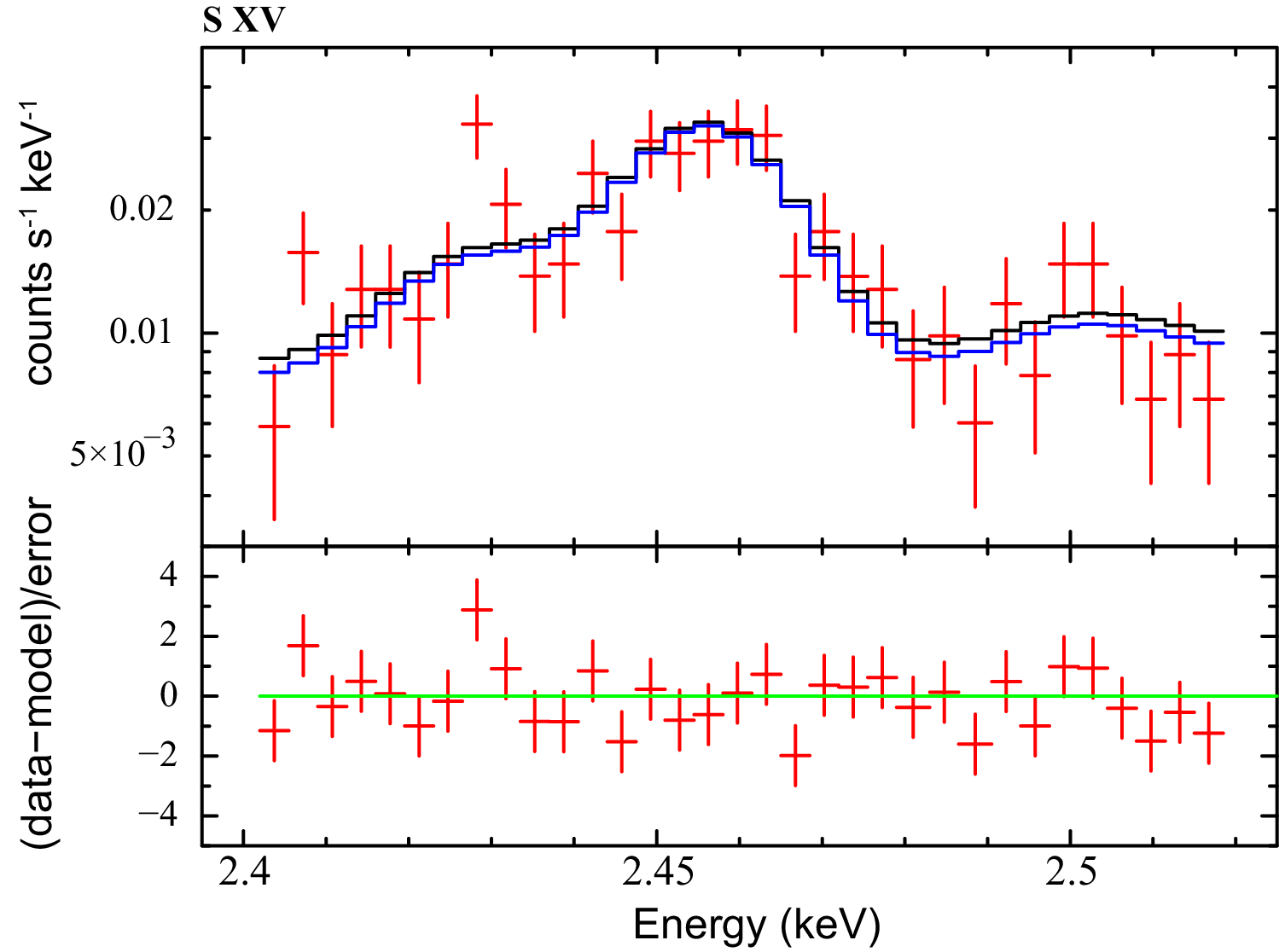}
    \includegraphics[width=0.48 \textwidth]{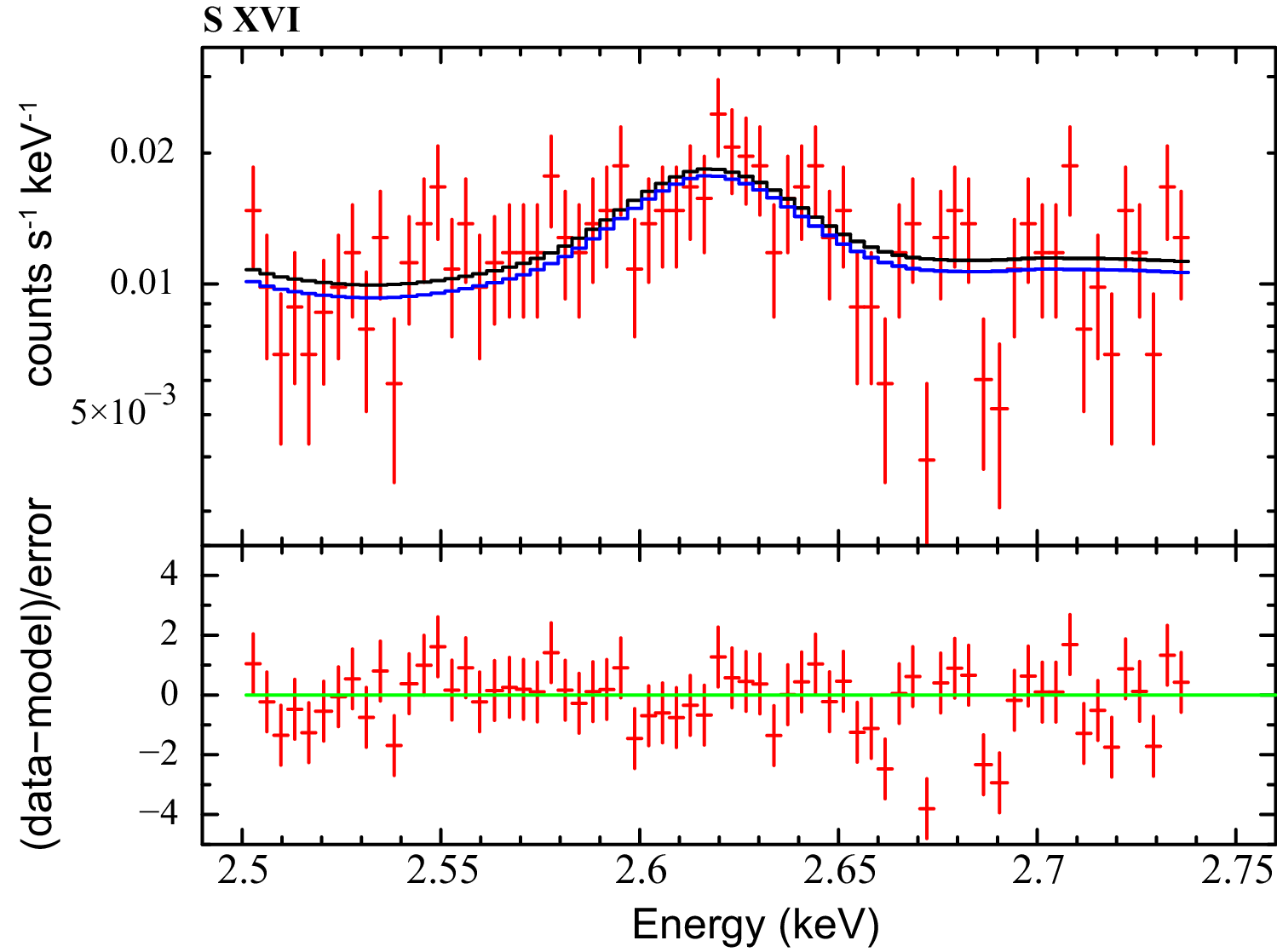}\\
    \includegraphics[width=0.48 \textwidth]{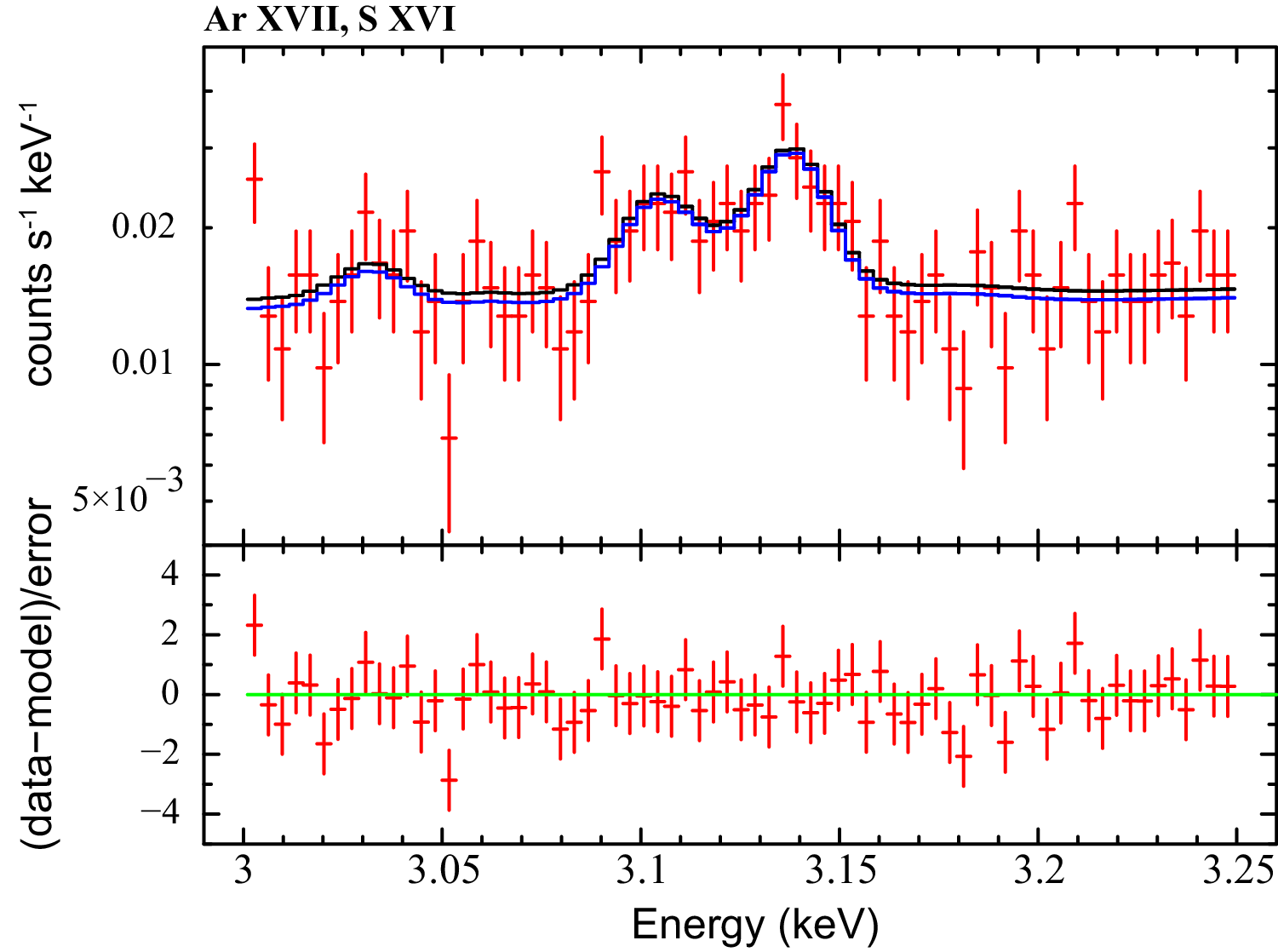}
    \includegraphics[width=0.48 \textwidth]{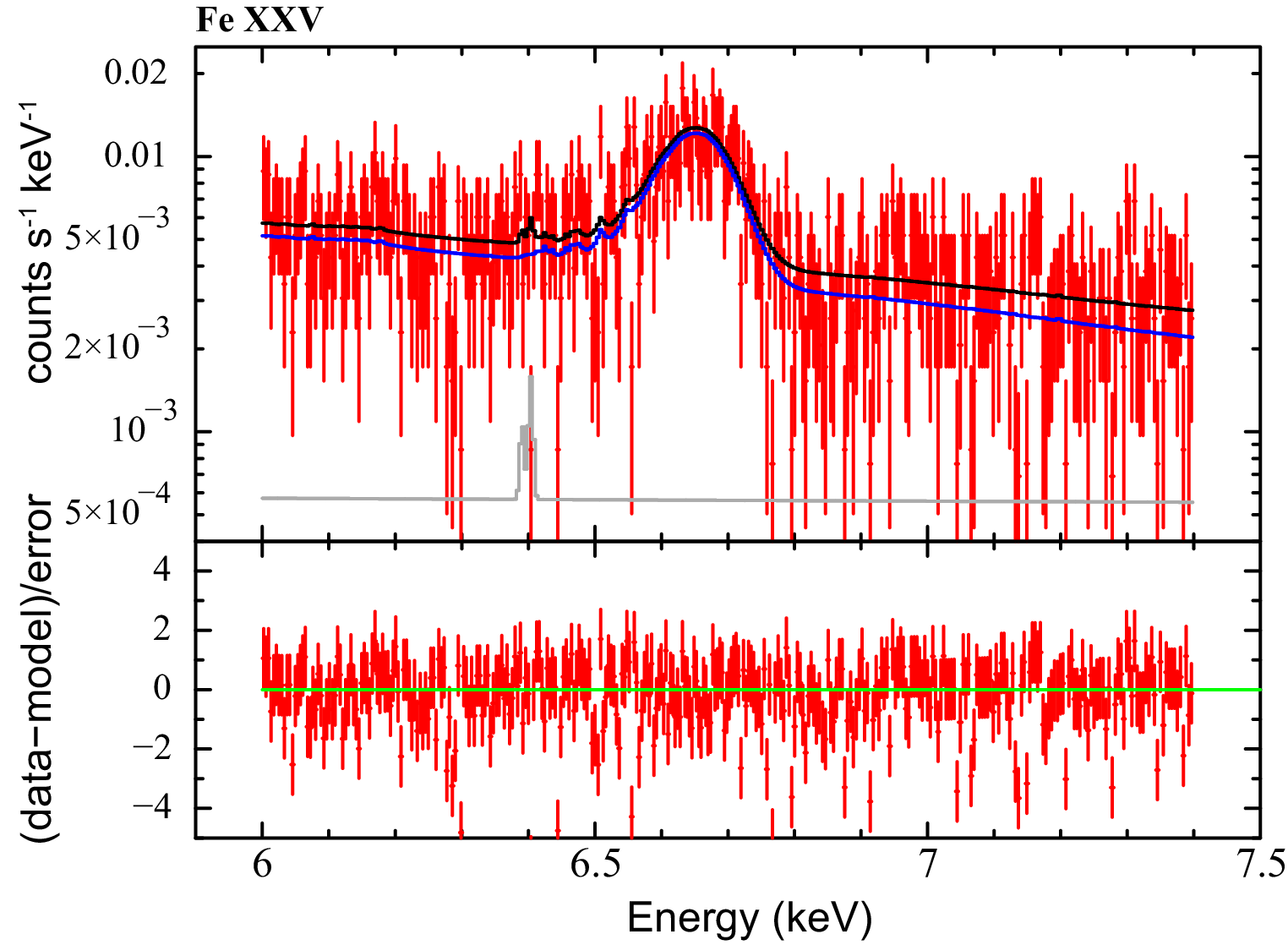}
    \end{center}
    \caption{The X-ray spectra around the Si XIII, Si XIV, S XV, S XVI, Ar XVII, and Fe XXV lines, from top left to bottom right, respectively. 
    The best fit models using {\it bvpshock} model with the parameters in Table \ref{tab:resolve_line_fit_param} are shown in blue, and its sum with the NXB model is shown in gray.
    The bottom panels represent the $\chi^2$ from the best fit models. 
    {Alt text: Six line graphs showing the data and best-fit models with residuals for each atomic line.} 
    }
    \label{fig:sn1987a_line_fit}
\end{figure*}
%============== Figure 6 ====================

Figure \ref{fig:sn1987a_line_fit} and Table \ref{tab:resolve_line_fit_param} provide a summary of the line fitting analyses. 
The Doppler shift $z$ exhibits significant errors that align with the intrinsic velocity, as previously seen in Section \ref{sec:data_analyses_spectrum:resolve_all}. 
The line widths for six lines fall within a Doppler velocity range of 800 -- 1700 km s$^{-1}$.

%=================================================================
\section{Discussion}
\label{sec:discussion}
%=================================================================
%------------------------------------------------
\subsection{Origin of Plasma Emission with Resolve}
\label{sec:discussion:origin}
%------------------------------------------------
Previous X-ray observations consistently require a multi-temperature plasma to characterize the X-ray spectrum of SN~1987A (e.g., \cite{1994A&A...281L..45B,1996A&A...312L...9H,2000ApJ...543L.149B,2002ApJ...567..314P,2004ApJ...610..275P,2005ApJ...628L.127Z,2005ApJ...634L..73P,2006ApJ...645..293Z,2006ApJ...646.1001P,2008ApJ...676L.131D,2009ApJ...703.1752R,2009ApJ...692.1190Z,2009PASJ...61..895S,2010A&A...515A...5S,2010MNRAS.407.1157Z,2011ApJ...733L..35P,2012A&A...548L...3M,2013ApJ...764...11H,2016ApJ...829...40F,2021ApJ...916...41S,2021ApJ...922..140R,2022A&A...661A..30M,2024ApJ...966..147R}, and references therein). 
There is an indication of contamination in the Resolve spectrum by a cooler component, hinted at in the low energy band but not in the higher two bands, as suggested in Figure \ref{fig:sn1987a_3band_cont} top.
However, the spectral characteristics across the three bands (Sections \ref{sec:data_analyses_spectrum:fe_band} and \ref{sec:data_analyses_spectrum:resolve_3band}) are consistent with each other at the 99\% confidence level. 
Therefore, we conclude that the Resolve spectrum in the 1.5 -- 10 keV range is well described by the single component using a plane-parallel shock plasma model {\it pshock} family (i.e., {\it vpshock} or {\it bvpshock} models in XSPEC), characterized by a temperature of $2.84_{-0.08}^{+0.09}$ keV and an ionization parameter of $2.64_{-0.45}^{+0.58} \times 10^{11}$ s cm$^{-3}$, as detailed in Section \ref{sec:data_analyses_spectrum:resolve_all}. 
This temperature measured by Resolve at 37.3 years post-explosion aligns closely with the trend from Chandra monitoring observations (see Fig. 3 in \cite{2024ApJ...966..147R}). 

\begin{figure}[htb]
    \begin{center}
    \includegraphics[width=0.48 \textwidth]{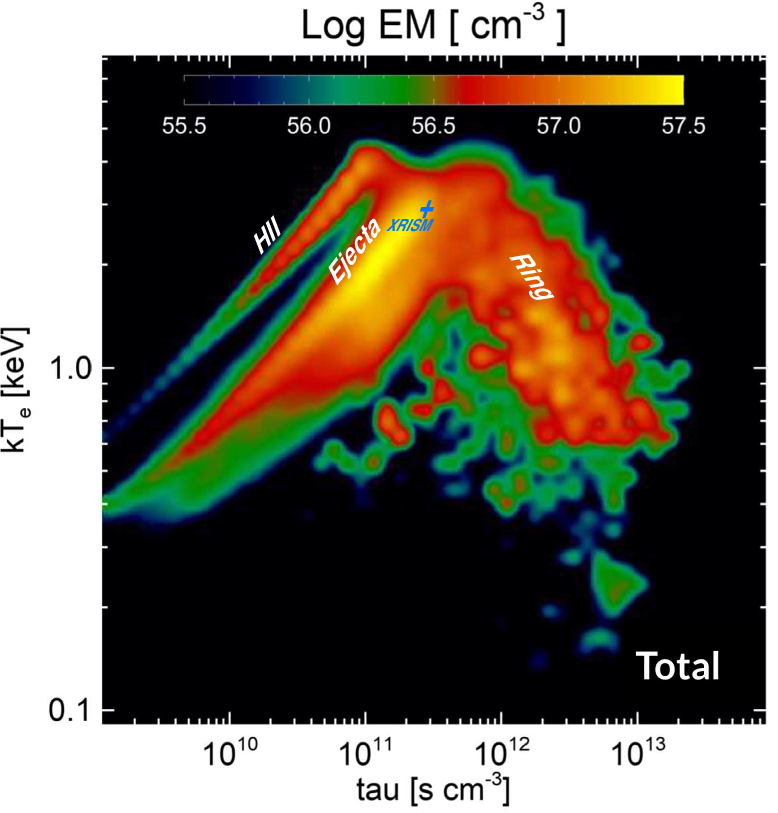}
    \end{center}
    \caption{The distribution of EM for X-ray emitting plasma originating from ejecta, H{\sc ii}, and the ring is shown as a function of $kT$ and $\tau$, copied from Figure 1 in \citet{2024ApJ...961L...9S}. 
    The observational result from Resolve, referred to as "pshock\_wide2" in Table \ref{tab:resolve_broad_fit_param_sun21}, is also illustrated.
    {Alt text: One image graph with color scale shown in two parameter spaces estimated from the simulation and the best-fit value from our observation.} 
    }
    \label{fig:sn1987a_kt_nt_plot}
\end{figure}

We consider our results to be indicative of a dominant contribution of shocked ejecta to the observed X-ray emission, as explained below.
Fig. \ref{fig:sn1987a_kt_nt_plot} shows the distribution of the plasma emission measure (EM) as a function of $kT$ and the ionization parameter $\tau$ predicted by dedicated state-of-the-art 3D MHD simulations of SN 1987A \citep{2020A&A...636A..22O,2024ApJ...961L...9S}, where we superimpose our best-fit values. 
Both $kT$ and $\tau_{\rm u}$ derived from the Resolve spectral analysis are consistent with the values predicted by the simulation for the ejecta component.
We warn the reader that the EM distribution presented in Fig. \ref{fig:sn1987a_kt_nt_plot} shows the plasma conditions \emph{at the source}, i.e. without taking into account the effects of the interstellar absorption and the instrumental response. 
The interstellar absorption around the blast wave is suggested in the soft X-ray observations \citep{2021ApJ...916...41S,2024ApJ...966..147R,2025ApJ...981...26S}, as well as by optical evolutions \citep{2015ApJ...806L..19F,2019ApJ...886..147L}, declining IR light curve \citep{2016AJ....151...62A}.
Since both interstellar absorption and the effects of closed GV (Section \ref{sec:observation}) suppress mainly the soft X-ray emission, the observed Resolve spectrum will be more biased toward the hard X-ray emitting portion of the EM distribution.
Indeed, we notice that the best-fit values resulting from the Resolve spectral analysis are slightly higher than those predicted for the ejecta, as expected, while they are clearly not consistent with those of the H{\sc ii} region and the Ring, whose density is taken into account in the 3D MHD simulations assuming to be located within a CSM structure characterized by a uniform density of 50 cm$^{-3}$ \citep{2020A&A...636A..22O,2024ApJ...961L...9S}.
The large line broadening that we measure also indicates that ejecta play an important role in shaping the line profile, as we discuss in detail in Section \ref{sec:discussion:doppler}.

%============== Figure 7 ====================
%============== Figure 7 ====================

\begin{figure}[htb]
    \begin{center}
    \includegraphics[width=0.48 \textwidth]{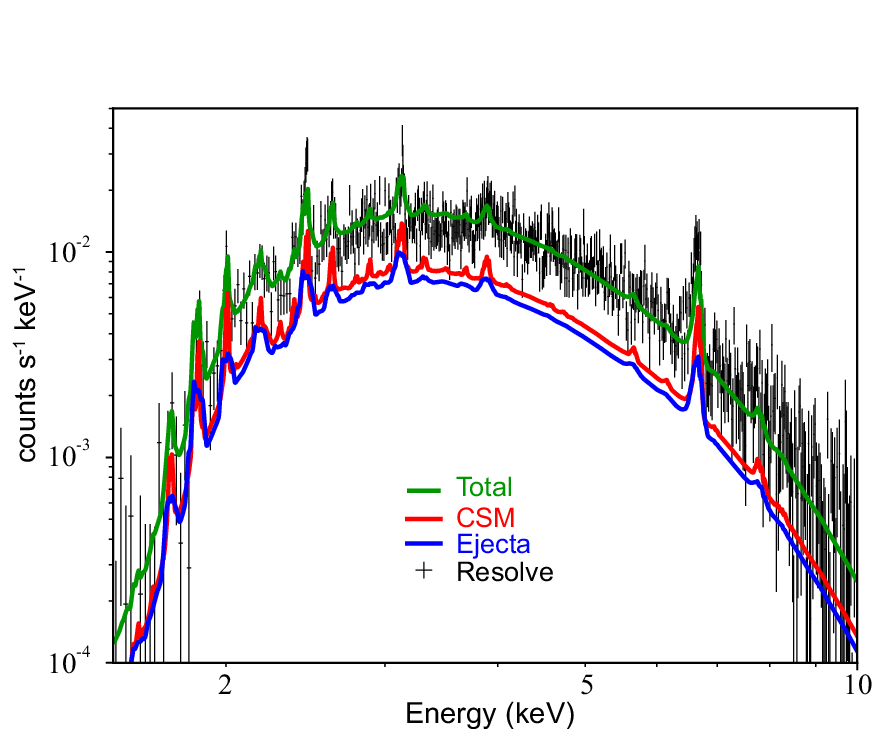}
    \end{center}
    \caption{The background-subtracted Resolve spectrum with the synthetic spectral models for ejecta and CSM for the year 2024, estimated using the MHD simulation from \citet{2020A&A...636A..22O}, are displayed in black, blue and red, respectively.
    Green line represents the sum of ejecta and CSM models.
    {Alt text: Single line graph showing both the data and the three models predicted by the 3D simulation.}
    }
    \label{fig:Resolve_spectrum_MHDsimulation}
\end{figure}

To better understand the respective contributions of the components to the X-ray spectrum, we synthesize the XRISM-Resolve spectrum of SN 1987A from the MHD simulation presented in \citet{2020A&A...636A..22O}. 
To this end, we follow the methodology described in \citet{2024ApJ...961L...9S}, including the effects of the closed GV condition \citep{2024RNAAS...8..156S}.
Figure \ref{fig:Resolve_spectrum_MHDsimulation} shows the comparison between the observed (black crosses) and synthetic (dashed black curve) XRISM-Resolve spectra of SN 1987A, where we highlighted in red and blue the contributions of circumstellar matter (CSM; H{\sc ii} plus ring) and ejecta, respectively. 
The synthetic spectrum agrees remarkably well with the data and self-consistently reproduces all the main spectral features without the need for performing ad-hoc fitting.
This agreement also suggests a strong contribution by the shocked ejecta to the observed X-ray emission.
An upcoming paper will provide a detailed summary comparing XRISM spectra with MHD simulations.

%============== Figure 8 ====================
%============== Figure 8 ====================

%------------------------------------------------
\subsection{Non metal-rich Ejecta Emission}
\label{sec:discussion:abundance}
%------------------------------------------------

%%% non metal rich (XRISM result)
As outlined in Section \ref{sec:discussion:origin}, the plasma characteristics derived using XRISM (see Tables \ref{tab:resolve_broad_fit_param_sun21} and \ref{tab:resolve_broad_fit_param_stm10_rav24}) along with the 3D MHD simulations by \citet{2020A&A...636A..22O} indicate a non-negligible contribution from shock-heated ejecta to the Resolve spectra.
Typically, we expect metal-rich emission from hot ejecta in supernova remnants. 
However, the plasma abundances measured with Resolve align with those of the LMC, as detailed in the same Tables \ref{tab:resolve_broad_fit_param_sun21} and \ref{tab:resolve_broad_fit_param_stm10_rav24}.
This finding implies that the X-ray emissions originate from 'non-metal-rich' shock-heated ejecta.

%%% Interpretation.
As predicted by \citet{2024ApJ...961L...9S} using 3D MHD simulations \citep{2020A&A...636A..22O} and as already pointed out by \citet{2024ApJ...966..147R} and \citet{2025ApJ...981...26S}, the outer envelope (or mantle) of the ejecta, expected to reflect the LMC's chemical composition, contributes to the X-ray emission in 2024.
Therefore, Resolve spectra clearly confirmed that the reverse shock has not reached the inner part of the metal-rich ejecta yet.
Continuous future X-ray monitoring of high-resolution spectra of SN~1987A will facilitate the transition of plasma emission from non-metal-rich to metal-rich ejecta \citep{2025arXiv250419896}.
Considering that the progenitor of SN~1987A is not a stripped-envelope SN where metal-rich ejecta expand unimpeded like in Cassiopeia A, we expect SN~1987A to have an extensive H-rich envelope of ejecta, thus resulting in a year-scale time span in the transition from non-metal-rich emission.

%------------------------------------------------
\subsection{Line Doppler broadening and shift}
\label{sec:discussion:doppler}
\begin{figure}[htb]
    \begin{center}
    \includegraphics[width=0.48 \textwidth]{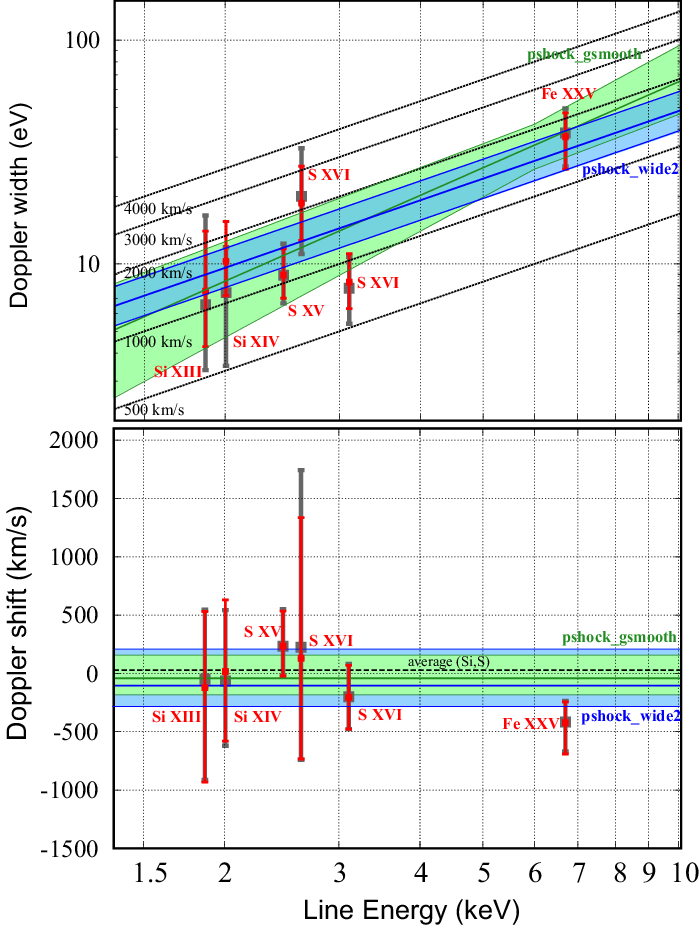}
    \end{center}
    \caption{The top and bottom panels illustrate the Doppler width and shift values, respectively, obtained through the line fitting in Section \ref{sec:data_analyses_line}, as functions of line energies. 
    The red and gray open squares, accompanied by 90\% error bars, indicate the outcomes of the line fittings in Table \ref{tab:resolve_line_fit_param} with abundance either fixed or variable, respectively. 
    Additionally, the broad band fitting results, labeled "pshock\_wide2" and "pshock\_gsmooth" in Table \ref{tab:resolve_broad_fit_param_sun21}, are also plotted with blue and green lines showing the 90\% confidence regions.
    On the bottom panel, the dashed line represents the average Doppler shift values for Si and S lines.
    {Alt text: Two line graphs plotting the analyses results.} 
    }
    \label{fig:sn1987a_doppler_plot}
\end{figure}

%------------------------------------------------
%% Summary of our analyses
Figure \ref{fig:sn1987a_doppler_plot} summarizes the Doppler widths of the atomic lines and the shift inferred from line fitting in Section \ref{sec:data_analyses_line}, overlayed on the results of the full-band fitting in Section \ref{sec:data_analyses_spectrum:resolve_all}.
The plots are shown as a function of the energy of atomic lines.
On the top panel of the figure, the result of Doppler broadening predicted by the "pshock\_wide2" model assumes the energy power $\alpha = 1$ (i.e., constant velocity among lines), as described in Section \ref{sec:data_analyses_spectrum:resolve_all}, and the plot by "pshock\_gsmooth" reflects the best-fit value of $\alpha = 1.25_{-0.37}^{+0.32}$.

%============== Figure 9 ====================
%============== Figure 9 ====================

%%% statement: bulk rather than thermal
In this section, we assume that the contribution from the 'non-metal-rich' shocked ejecta dominates the Resolve spectrum, as concluded in Sections \ref{sec:discussion:origin} and \ref{sec:discussion:abundance}.
The Doppler widths of the atomic lines, as indicated in Tables \ref{tab:resolve_broad_fit_param_sun21} and \ref{tab:resolve_broad_fit_param_stm10_rav24}, ranged between $1,500$ and $1,800$ km s$^{-1}$.
Since the ejecta abundance aligns with the LMC value (Section \ref{sec:discussion:abundance}), the electron and ion temperatures are expected to have reached equilibrium at $\tau \sim 10^{10 - 11}$ cm$^{-3}$ s, even for Fe, whereas higher abundances at $10^5$ solar require waiting until $\tau \sim 10^{13}$ due to the significant mass difference with protons.
Consequently, the Doppler broadening of $1,500$ -- $1,800$ km s$^{-1}$ observed by Resolve considerably exceeds the thermal broadening width at approximately $kT \sim 2.8$ keV. 
Furthermore, as \citet{2021ApJ...916...41S} noted, the energy power $\alpha > 1$ implies that heavier atomic lines exhibit higher velocities, although the significance is low statistically.
These facts support the notion that the Doppler broadening of the lines seen in the Resolve spectra is due to the kinematic expansion structure of SN~1987A.
In addition, the thermal broadening measured by \citet{2019NatAs...3..236M} during the CSM dominant phase would correspond to less than 450 km s$^{-1}$, which is well below the broadening measured in our spectral analysis during the current ejecta-emitting phase. 
This confirms again that the line broadening is determined by the large bulk motion of the X-ray emitting plasma.
Moreover, the energy dependency found with XRISM ($\alpha = 1.25_{-0.37}^{+0.32}$) approaches to the unity (i.e., constant velocity over the atomic lines) from that reported by \citet{2021ApJ...916...41S} during the CSM dominant phase ($\alpha\sim2$), indicating that the contribution from kinematic Doppler broadening in the ejecta to the line widths is increasing at the time of the XRISM observation.

%% Velocity in Radio/NIR/X-ray
According to the morphological analyses of the evolving ER images in the soft X-rays \citep{2016ApJ...829...40F,2024ApJ...966..147R,2025ApJ...981...26S}, 9 GHz radio \citep{2013ApJ...777..131N,2018ApJ...867...65C}, and mid-infrared (MIR; \cite{2022MNRAS.517.4327M}) bands, the physical expansion velocities are initially at $3,000$-- $4,000$ km s$^{-1}$ before $\sim 7,000$ days after the explosion, and re-acceleration happens prior to reaching the ER; however, only the velocities in the soft X-ray band had decreased to about $2,000$ km s$^{-1}$ until recently.
The Doppler widths seen in the Resolve spectra at $1,500$ -- $1,800$ km s$^{-1}$ align approximately with these recent values, suggesting that the emission from materials with lower velocities than the CSM located beyond the ER should be dominant in the Resolve spectra.
Small diversity of the velocities across different wavelengths originates from distinct radiation mechanisms that investigate various material states. 
For example, recent evolution of near-infrared (NIR) images is rather slower (at approximately $690$ km s$^{-1}$; \cite{2023A&A...675A.166K}) than those in other wavelengths because NIR/optical images represent stable hot-spots on ER.
Similarly, synchrotron radiation in the radio band originates from behind the shock, while thermal X-ray emissions explore high emission-measure (EM) regions. 
Notably, the forward shock has left the ER and is now traveling through a low-density medium with a high velocity, but the thermal X-rays are insensitive to such regions because they probe dense CSM like the ER and dense clumps.
Consequently, atomic lines from the CSM are expected to appear narrow in the X-ray spectra.
In contrast, the X-ray emitting ejecta are heated by the reverse shock generated via the impact with the ER, which has an anisotropic velocity structure (e.g., a bi-polar flow seen in the JWST NIR image by \cite{2023ApJ...949L..27L}), and the lines from the ejecta are expected to be broad in X-rays, as observed with Resolve.

\begin{figure*}[htb]
    \begin{center}
    \includegraphics[width=0.95 \textwidth]{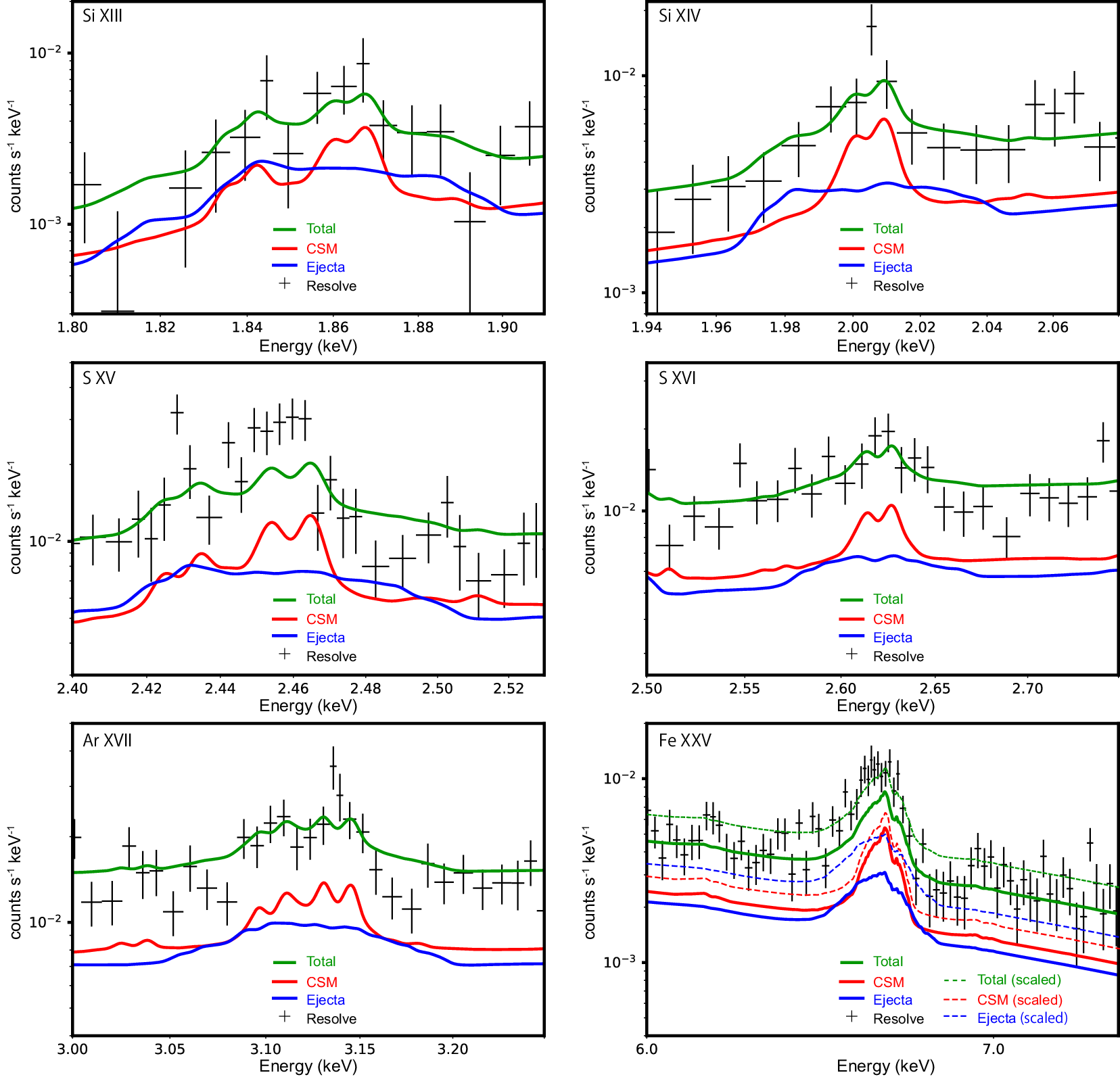}
    \end{center}
    \caption{Same as Fig. \ref{fig:Resolve_spectrum_MHDsimulation}, but close-up around Si XIII, Si XIV, S XV, S XVI, Ar XVII, and Fe XXV lines. 
    Black, blue, red, and green lines represent the background-subtracted Resolve spectrum, ejecta model, CSM model, and sum of ejecta and CSM models.
    For Fe XXV line, blue, red, and green dashed line represents the models of ejecta, CSM, and sum of them, respectively, scaled by 60\% and 20\% for ejecta and CSM, respectively. 
    {Alt text: Six line graphs showing both the data and the models predicted by the 3D simulation for each line.} 
    }
    \label{fig:Resolve_spectrum_MHDsimulation_zoom}
\end{figure*}

%% Comparison with 3D simulation
As seen in Fig. \ref{fig:Resolve_spectrum_MHDsimulation} discussed in Section \ref{sec:discussion:origin}, the 3D MHD simulations of SN 1987A \citep{2020A&A...636A..22O,2024ApJ...961L...9S} reproduce our Resolve spectrum well.
To gain a deeper understanding of the physical origin of the observed line broadening, Fig.\ref{fig:Resolve_spectrum_MHDsimulation_zoom} shows different close-up views of the observed and synthetic spectra focused on specific emission lines. 
As anticipated in \citet{2024ApJ...961L...9S}, we observe, on average, a predominance of the CSM contribution only at the center of the emission lines, with a striking contribution of the ejecta at the wings, even though the 3D MHD simulations encompassed contributions from the entire CSM, incorporating not only the ER but also the dense clumps, the uniform ring component, and the HII region.
Modeling the line widths suggests that the observed line profiles can be explained with a weighted average of Doppler broadening between the CSM and ejecta, where the broad ejecta component is dominant.
For this reason, the velocities we retrieve from our spectral analysis are slightly smaller than those predicted for the ejecta by the MHD model (which are of the order of 3000 km s$^{-1}$). 
Nevertheless, the line widths we measure are definitely much larger than those expected for the CSM. 
We conclude that the enhanced line widths observed by XRISM-Resolve provide the first clear evidence of X-ray emission from shocked ejecta in SN 1987A. 

%% Fe line; obs vs sim
The close-up view of the Fe line in Fig.\ref{fig:Resolve_spectrum_MHDsimulation_zoom} provides some interesting hints; we notice that the model under-reproduces both the line flux and its width. 
This is indicative of both i) a larger contribution from the Fe-rich plasma with respect to that predicted by the model and ii) a larger velocity of the Fe-rich material as an additional contribution.
In particular, we notice that the differences between the synthetic and actual line profiles are enhanced in the low-energy wing of the line.
New dedicated MHD simulations are expected to address this issue (Orlando et al.~ in preparation). 
A detailed comparison between the MHD model and the actual observation is beyond the scope of this article and will be addressed in a forthcoming publication (Sapienza et al. in preparation), but we speculate that this may be indicative of red-shifted Fe-rich ejecta in \citet{2023ApJ...949L..27L}. 

%============== Figure 10 ====================
\subsection{Search for Emission from a Pulsar Wind Nebula}
\label{sec:discussion:pwn}
%------------------------------------------------
%%  PWN Introduction
The emission of the pulsar wind nebula (PWN) in SN~1987A system is suggested and detailed by \citet{2021ApJ...908L..45G} and \citet{2022ApJ...931..132G}. 
We investigated the potential presence of emission from a possible PWN using the Resolve spectrum. 
To model the PWN emission described by \citet{2022ApJ...931..132G}, the {\it vphabs*powerlaw} model was added to the best fit model for the thermal emission. 
In this model, the column density for {\it vphabs} was set to $10^{23}$ atoms cm$^{-2}$, the abundances of absorbing elements were aligned with those of the "pshock\_wide2" model from Table \ref{tab:resolve_broad_fit_param_sun21}, and the photon index for {\it powerlaw} was fixed at 2.8. 
As illustrated by the orange line in Figure \ref{fig:sn1987a_broad_spectra}, any PWN emission is approximately an order of magnitude weaker than the signal in the Resolve energy band. 
Although a neutral Fe-K absorption edge at around 7.1 keV could be anticipated in the high-resolution spectrum with Resolve, no notable feature was detected.

%% Upper limit
To assess the upper limit of the PWN emission using Resolve, we performed a spectral fitting in the 1.5--10 keV range. 
The photon index and abundance pattern in the PWN model were fixed, while the parameters $kT, \tau_{\rm u}, z$, and Doppler velocity of the {\it bvpshock} model were freely varied. 
As a result, we estimate that the 90\% upper limit flux of the PWN component is $1.0 \times 10^{-13}$ erg cm$^{-2}$ s$^{-1}$ in the 10 to 20 keV band and $4.3 \times 10^{-13}$ erg cm$^{-2}$ s$^{-1}$ in the 2 to 10 keV band. 
Assuming a distance $d = 51.4$ kpc, the values correspond to upper limit luminosities of $3.2 \times 10^{34}$ erg s$^{-1}$ and $1.4 \times 10^{35}$ erg s$^{-1}$ in the respective bands, aligned with the luminosity values reported by \citet{2022ApJ...931..132G}.

%------------------------------------------------
\subsection{Search for $^{44}$Sc K line}
\label{sec:discussion:44sc}
%------------------------------------------------
%%%%%%% 44Ti Introduction %%%%%%%%%
In the context of nucleosynthesis research on stellar evolution leading up to core-collapse supernovae (SNe), $^{44}$Ti is produced at the mass cut region, located at the interface between the innermost ejecta and the prospective compact object (\cite{1996ApJ...464..332T}; see also \cite{1998PASP..110..637D}), and is an alpha-rich freeze-out product, which happens when the density drops rapidly and not all $^{4}$He particles are used to build up completely to $^{56}$Ni.
Hence, observational measurements on the initial mass of $^{44}$Ti (denoted as $M_{\rm 44}$) are crucial for understanding the core-collapse explosion process and the accompanying nucleosynthesis. 
The $^{44}$Ti nuclei undergo decay into $^{44}$Sc and $^{44}$Ca as, 
\begin{equation}
    ^{44}{\rm Ti} \xrightarrow{(EC)}{} ^{44}{\rm Sc} \xrightarrow{(\beta^+)}{} ^{44}{\rm Ca}, \nonumber
\end{equation}
where EC and $\beta+$ represent the electron capture and $\beta+$ decays, respectively.
This decay process produces a number of nuclear decay lines from excited $^{44}$Sc at 67.87 keV ($W_{\rm 68}=93.0$\% of decays) and 78.32 keV (96.4\% of decays) with a characteristic decay time of $t_{\rm 44}=85.0$ years, and from excited $^{44}$Ca at 1,157 keV (99.9\% of decay) \citep{1998PASP..110..637D,2006PhRvC..74f5803A} with a decay time of 5.4 hours. 
The latter time scale is much faster than $t_{\rm 44}$, and thus the $^{44}$Ti decay determines the overall time scale.
The $\beta^+$ decay (99.3\%) also results in a subsequent electron positron annihilation producing 511 keV radiation, and it leaves a K-shell electron hole in the $^{44}$Sc atom, which results in a fluorescence line at 4.09 keV ($W_{\rm Sc}=17.4$\% of decays).
The nuclear decay lines associated with the decay of $^{44}$Sc have only solidly been measured in two SNRs,
Cassiopeia A using CGRO/Comptel \citep{1994A&A...284L...1I}, BeppoSAX \citep{2001ApJ...560L..79V}, INTEGRAL \citep{2006ApJ...647L..41R} and NuStar \citep{2014Natur.506..339G}, and SN1987A by INTEGRAL \citep{2012Natur.490..373G} and NuStar \citep{2015Sci...348..670B}. 
Here, we focus on the accompanying $^{44}$Sc fluorescent line at 4.09 keV and search for potential line features in high-resolution spectra using Resolve.

%============== Figure 11 ====================
\begin{figure}[htb]
    \begin{center}
    \includegraphics[width=0.48 \textwidth]{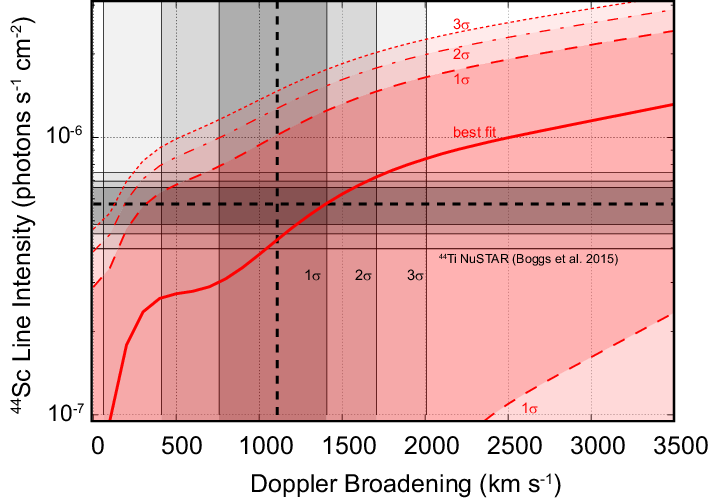}
    \end{center}
    \caption{The Results for the $^{44}$Sc-K line search with Resolve spectrum are illustrated in red.
    The thick, dashed, dot-dashed, and dotted lines correspond to the best fit, 1$\sigma$, 2$\sigma$, and 3$\sigma$ values, respectively, of the line intensities calculated by each Doppler broadening width. 
    The expectation based on $^{44}$Ti observations from NuSTAR \citep{2015Sci...348..670B} is represented by a black thick dashed line, with 1$\sigma$, 2$\sigma$, and 3$\sigma$ errors depicted in gray. 
    {Alt text: Single line graph showing the fitting result in contours and the prediction by previous observation.} 
    }
    \label{fig:sn1987a_44sc_search}
\end{figure}
%============== Figure 11 ====================

%%%%%%% XRISM Results %%%%%%%%%
Within the Resolve spectrum, no prominent line feature is noted at 4.09 keV. 
To determine the upper limit of the line intensity, we fitted the spectrum between 3.5 and 4.5 keV using a power law model combined with a Gaussian function. 
It should be noted that Resolve data were accumulated from all pixels except for calibration pixel \#12 (Section \ref{sec:data_reduction_validation:process}), and we did not perform this search constraining to image center pixels due to the moderate point-spread function of the mirror.
The line center remained fixed at 4.09 keV and was adjusted by a Doppler shift of -1,000 km s$^{-1}$, consistent with the findings of \citet{2015Sci...348..670B} using NuSTAR, under the assumption that the bulk of $^{44}$Ti and $^{44}$Sc exists still in the un-shocked ejecta and is not highly ionized.
Note that this Doppler shift value is the sum of intrinsic velocity of SN~1987A and the recession velocity of SN~1987A at 286.7 km s$^{-1}$ \citep{1987ICRC....1..164C,1995Ap&SS.233...75M}. We also tried the case with the recession velocity only, but the following result does not change.
It should be noted that $^{44}$Ti in Cassiopeia A is predicted to be shock-ionized \citep{2015Sci...348..670B}, however, the scenario for SN~1987A is expected to differ, as indicated by the differing Doppler shifts: $^{44}$Ti exhibits a blueshift \citep{2015Sci...348..670B}, while the shock-heated plasma represents different shifts as seen in Section \ref{sec:discussion:doppler}.
As a result, stricter upper limits of Gaussian normalizations were derived for narrower models, so we calculated the upper limits for each Doppler broadening velocity. 
Figure \ref{fig:sn1987a_44sc_search} illustrates the fitting results on Gaussian normalizations at 1, 2, and 3 $\sigma$ levels.
%The results do not change so much if we adopt the Doppler shift only by the recession velocity of SN~1987A at 286.7 km s$^{-1}$ \citep{1987ICRC....1..164C,1995Ap&SS.233...75M}.

%%%%%%% Interpretation %%%%%%%%%
The intensity of the 67.87 keV line of $^{44}$Ti, as measured by NuSTAR \citep{2015Sci...348..670B}, is $F_{\rm Ti} = (3.4\pm0.7) \times 10^{-6}$ photons cm$^{-2}$ s$^{-1}$, observed at $t_{\rm NuSTAR}=$ 26.5 years after explosion. 
This value can be transformed into the line flux of $^{44}$Sc for the epoch $t_{\rm XRISM} = 37.3$ years with XRISM via the relation 
\begin{eqnarray}
    F_{\rm Sc} &=&  
    \frac{W_{\rm Sc}}{W_{\rm 68}} \exp\left(\frac{-(t_{\rm XRISM}-t_{\rm NuSTAR})}{t_{\rm 44}}\right) F_{\rm Ti} \nonumber \\
    &\sim& 
    (0.58\pm0.12) \times 10^{-6} {\rm ~photons ~cm}^{-2} {\rm s}^{-1}. 
\end{eqnarray}
This line intensity and the $^{44}$Ti broadening by NuSTAR \citep{2015Sci...348..670B} are also plotted in Fig.\ \ref{fig:sn1987a_44sc_search}.
Thus, the upper limit of the $^{44}$ Sc line measured with Resolve is in agreement with the NuSTAR observation of $^{44}$ Ti within the confidence level of 1.0 $\sigma$.
The $1\sigma$-, $2\sigma$- and $3\sigma$-upper limits for the intensity of the $^{44}$Sc line, matched with the NuSTAR Doppler broadening for $^{44}$Ti, were $1.0, 1.3,$ and $1.5 \times 10^{-6}$ photons cm$^{-2}$ s$^{-1}$, respectively.
Consequently, the 1$\sigma$ upper limit on $M_{\rm 44}$ with XRISM is,
\begin{eqnarray}
    M_{\rm 44} &=&
    4 \pi \left(\frac{d}{\rm 51.4 ~kpc}\right)^2 F_{\rm Sc} 44m_{\rm p}t_{\rm 44} \exp\left(\frac{t_{\rm XRISM}}{t_{\rm 44}}\right) 
    \nonumber\\
    &<& 2.6 \times 10^{-4} M_{\odot},
\end{eqnarray}
where $m_{\rm p}$ and $M_{\odot}$ are the proton mass and the solar mass, respectively.
Considering that the PWN emission is highly absorbed \citep{2022ApJ...931..132G} as discussed in Section \ref{sec:discussion:pwn}, the $^{44}$Sc K emission could be affected by dusty ejecta. 
If we apply the same photo absorption {\it vphabs} in Section \ref{sec:discussion:pwn}, the flux of the $^{44}$Sc K emission is expected to become 0.67 times the original, and the upper limit of $M_{\rm 44}$ becomes $1.7 \times 10^{-4} M_{\odot}$.
In this context, Compton scattering causes negligible absorption for the gas with $10^{23}$ atoms cm$^{-2}$, resulting in approximately a 2\% decrease with an energy line shift of less than 60 eV.
Therefore, the XRISM-derived upper limit for the yield of $^{44}$ Ti is consistent with previous observations using NuStar \citep{2015Sci...348..670B}, INTEGRAL \citep{2012Natur.490..373G}, and Chandra \citep{2006ApJ...651.1019L}, and theoretical expectations using the simulations in \citet{2020A&A...636A..22O} and \citet{2020ApJ...888..111O}.

%============================================
\section{Conclusion}
\label{sec:conclusion}
%============================================
We performed observations of SN~1987A using XRISM, with a total exposure of 290.5 ksec in June 2024, and succeeded in acquiring high-resolution spectra from shock-heated plasmas in SN~1987A utilizing the onboard Resolve instrument. 
The spectra, ranging from 1.7 to 10 keV, are accurately described by a single-component, plane-parallel shock plasma model with a $kT=2.84_{-0.08}^{+0.09}$ keV and $\tau=2.64_{-0.45}^{+0.58} \times 10^{11}$ s cm$^{-3}$. 
Additionally, the 3-D magneto-hydrodynamic simulation presented by \citet{2020A&A...636A..22O} replicates the spectrum well, indicating that the X-ray emissions detected by XRISM Resolve are primarily contributed by shocked ejecta. 
A prominent Fe-K emission line is noted at $6.644_{-0.008}^{+0.011}$ keV, featuring a line width of $58.2_{-6.2}^{+9.0}$ eV and a flux of $7.35_{-0.85}^{+1.00} \times 10^{-14}$ erg cm$^{-2}$ s$^{-1}$, in agreement with previous X-ray studies \citep{2021ApJ...916...41S, 2024ApJ...966..147R,2025ApJ...981...26S}. 
The spectrum from Resolve also shows atomic lines of Si, S, and Ar.
The Doppler widths of these lines as well as the Fe line correspond to the velocities between 1,500 and 1,700 km s$^{-1}$, which exceed the thermal broadening expected from non-metal-rich shock-heated ejecta. 
This fact indicates that the Doppler broadening observed in the Resolve spectra arises from the kinetic expansion structure of SN~1987A. 

Beyond the plasma diagnostics above, we investigated signs of a neutron star in SN~1987A. 
A 90\% upper limit was established for non-thermal emissions from a PWN at $4.3 \times 10^{-13}$ erg cm$^{-2}$ s$^{-1}$ in the 2 -- 10 keV energy range, aligning with NuSTAR's results \citep{2022ApJ...931..132G}. 
In addition, our search for the $^{44}$Sc K line yielded a $1\sigma$ upper limit of $1.0 \times 10^{-6}$ photons cm$^{-2}$ s$^{-1}$, which corresponds to an original $^{44}$Ti mass of around $2 \times 10^{-4} M_{\odot}$, consistent with previous X-ray and soft gamma-ray studies \citep{2015Sci...348..670B,2012Natur.490..373G,2006ApJ...651.1019L}.

%=================================================================
% Acknowledgement
%      note: acknowledgement should be placed at end of main text.(NOT after the Appendix.)
%=================================================================
\begin{ack}
The authors thank all members of the XRISM team, especially for their continuous efforts to operate the spacecraft and maintain instruments onboard.
%This work was supported by the JSPS Core-to-Core Program (grant number: JPJSCCA20220002), the Japan Society for the Promotion of Science Grants-in-Aid for Scientific Research (KAKENHI) Grant Number JP20K04009 (YT), JP23K20850 (KM), JP21H01095 (KM), JP23H01211 (AB), JP23K25907 (AB), JP24K17105 (YK), JP24K00677 (JS), and by NASA XRISM grant numbers 80NSSC23K1656 (PP) and NASA CXC contract number NAS8-03060 (PP).
This work was supported by the JSPS Core-to-Core Program (grant number: JPJSCCA20220002), JSPS KAKENHI grant numbers JP20K04009 (YT), JP23K20850 (KM), JP21H01095 (KM), JP23H01211 (AB), JP23K25907 (AB), JP24K17105 (YK), JP24K00677 (MN), JP21K03615 (MN), JP22H00158 (YF), JP22H01268 (YF), JP22K03624 (YF), JP23H04899 (YF), JP21K13963 (KH), JP24K00638 (KH), JP21K13958 (MM), JP24H00253 (HN),  JP20K14491 (KN), JP23H00151 (KN), JP19K21884 (HN), JP20H01947 (HN), JP20KK0071 (HN), JP23K20239 (HN), JP24K00672 (HN), JP24K17104 (SO), JP24K17093 (HS), JP21H04493 (TGT), JP20H01946 (UY), JP23K13154 (SY), JP19K14762 (MS), JP23K03459 (MS), JP20H05857 (NU), JP23K03454 (RF), JP23K22548 (YM), and NASA grant numbers 80NSSC18K0988 (PP), 80NSSC23K1656 (PP), 80NSSC20K0733 (EB), 80NSSC24K1148 (EB), 80NSSC24K1774 (EB), 80NSSC18K0978 (LC), 80NSSC20K0883 (LC), 80NSSC25K7064 (LC), 80NSSC20K0737 (EDM), 80NSSC24K0678 (EDM), 80NSSC18K1684 (IZ), and 80NNSC22K1922 (DM).
The material is based upon work supported by NASA under award number 80GSFC21M0002 (AO, KP, KH, TH, FM, RB, ML, KM).
%%%
P.P. acknowledges support from NASA CXC contract NAS8-0360.
M.M. acknowledges support from Yamada Science Foundation.
L.C. acknowledges support from NSF award 2205918.
C.D. acknowledges support from STFC through grant ST/T000244/1.
L.G. acknowledges financial support from Canadian Space Agency grant 18XARMSTMA.
M.S. acknowledges the support by the RIKEN Pioneering Project Evolution of Matter in the Universe (r-EMU) and Rikkyo University Special Fund for Research (Rikkyo SFR).
AT acknowledges the support by the Kagoshima University postdoctoral research program (KU-DREAM).
SU acknowledges support by the Program for Forming Japan's Peak Research Universities (J-PEAKS).
SY acknowledges support by the RIKEN SPDR Program.
IZ acknowledges partial support from the Alfred P. Sloan Foundation through the Sloan Research Fellowship.
Part of this work was performed under the auspices of the U.S. Department of Energy by Lawrence Livermore National Laboratory under Contract DE-AC52-07NA27344.
The material is based on work supported by the Strategic Research Center of Saitama University.
%%%
V.S. and M.M. acknowledge financial contributions from the PRIN MUR “Life, death and after-death of massive stars: reconstructing the path from the pre-supernova evolution to the supernova remnant” funded by the European Union - Next Generation EU.
%%%%
The authors appreciate the insightful comments and suggestions provided by the anonymous referee.
\end{ack}

\end{document}